\documentclass[aps,pre,reprint,superscriptaddress,floatfix]{revtex4-2}

\usepackage{amsfonts}
\usepackage{amssymb}
\usepackage{amsmath}
\usepackage{dcolumn}  
\usepackage[T1]{fontenc}
\usepackage{silence}
\usepackage{patches}
\usepackage{hyperref}
\usepackage{cleveref}
\usepackage{makecell}
\usepackage{xcolor}
\definecolor{myred}{RGB}{255,0,0}
\definecolor{myblue}{RGB}{0,0,255}
\definecolor{mygreen}{RGB}{0,255,0}
\usepackage{siunitx}

\begin{document}

\title{Simulations of Three-dimensional Nematic Guidance of Microswimmers}

\author{Zeyang Mou}
\affiliation{
Department of Physics, Hong Kong University of Science and Technology, Clear Water Bay, Kowloon, Hong Kong SAR
}

\author{Yuan Li}
\affiliation{
Fujian Provincial Key Laboratory for Soft Functional Materials Research, Research Institute for Biomimetics and Soft Matter, Department of Physics, Xiamen University, Xiamen, Fujian 361005, China}

\author{Zhihong You}
\affiliation{
Fujian Provincial Key Laboratory for Soft Functional Materials Research, Research Institute for Biomimetics and Soft Matter, Department of Physics, Xiamen University, Xiamen, Fujian 361005, China}

\author{Rui Zhang}
\email[]{ruizhang@ust.hk (R.Z)}
\affiliation{
Department of Physics, Hong Kong University of Science and Technology, Clear Water Bay, Kowloon, Hong Kong SAR
}

\date{\today}
% Length: within 3,500 words;
% Abstract:About 5% of article length & < 500 words; 
\begin{abstract}
%Introduction to Dynamics Control
It has been shown that an anisotropic liquid crystalline (LC) environment can be used to guide the self-propulsion dynamics of dispersed microswimmers, such as bacteria. This type of composite system is named ``living nematic''. In the dilute limit, bacteria are found to mainly follow the local director field. Beyond the dilute limit, however, they exhibit novel dynamical behaviors, from swirling around a spiral $+1$ defect pattern to forming undulating waves, and to active turbulence. Our current knowledge of how these different behaviors emerge at different population densities remains limited.
%Active units swimming in a liquid crystalline environment are demonstrated to effectively control their dynamics. By varying the population of microswimmers, their behavior can transform from strictly following a predesigned orientation to unidirectional collective propagation, circulation, or undulation. 
% Numerical Method Development
Here we develop a hybrid method to simulate the dynamics of microswimmers dispersed in a nematic LC. Specifically, we model the microswimmers as active Brownian particles whose dynamics are coupled to a hydrodynamic model of nematic LCs to describe the evolution of the flow field and the LC structure.
%the Brownian-like dynamics of swimmers coupled with director and flow fields. This interaction allows the superposition of active flows from individual microswimmers to modify the flow field, which subsequently distorts the director field.
% Validation through Experimentation:
Our method is validated across a wide range of microswimmer populations by comparing to existing quasi-two-dimensional (2D) experiments, including undulated swirling around a spiral $+1$ defect pattern and stabilized undulated jets on a periodic C-pattern.
%detailed comparisons with previous experimental results, including the individual to collective and stabilized undulation of bacteria in a spiral vortex or periodic splay-bend patterned thin cell, the accumulation of bacteria in the cores of $+1/2$ topological defects and depletion of bacteria in the cores of $-1/2$ defects. 
% 3D System Application
We further extend our method to three-dimensional (3D) systems by examining loop-defect dynamics. We find that the morphodynamics and destiny of a loop defect not only depend on the activity (self-propulsion velocity), effective size, and the initial distribution of the swimmers, but also rely on its winding profile. For a wedge-twist loop defect, its dynamics are mainly determined by the position and orientation of the $+1/2$ wedge. For a pure-twist loop defect, radial twist windings play a similar role as the $+1/2$ wedge in the wedge-twist loop defect, while other windings can engender out-of-plane active flows to buckle the pure-twist loop.
Finally, we consider the stochastic reversals of the self-propulsion direction of the microswimmers. By varying the characteristic reversal time, we predict that microswimmers do not necessarily tend to accumulate on splay regions.
Taken together, our hybrid method provides a faithful tool to explain and guide the experiments of living nematics in both 2D and 3D, sheds light on the interplay between microswimmer distribution and defect dynamics, and unravels the design principles of using LCs to control active matter.
%, proposing scenarios with different defect loops. Microswimmer density changes are shown to depend on self-propulsion ability, pairwise interactions, direction reversal time, and coupling with both local and bulk defect loop profiles. Regions of higher bacterial density consistently align with areas of high curvature and $+1/2$ wedge-type defects, while depletion occurs around segments of lower curvature and $-1/2$ wedge-type defects, regardless of loop type or boundary conditions. 
%Furthermore, swimmer migration influences defect loop dynamics by either promoting or inhibiting elongation (and potential breakage), compression, and lifting, while also driving locomotion and profile transformations under nonuniform initial distributions. 
% Implications and Applications
%These findings provide new strategies for directing microswimmers behavior, with potential applications in material transport and targeted delivery in anisotropic media, expanding the toolbox for manipulating active matter.
\end{abstract}


\maketitle

\section{Introduction}
% version 1
% para 1: microswimmer --> control collective behaviour --> Bacterial dispersions in lyotropic chromonic liquid crystals
% para 2: in this paper

% version 2
% para 1: Active materials --> collective motion --> (active) force dipoles can destabilize their nematic order via hydrodynamic instabilities 
% para 2: (control) bacteria swimming in nematics --> 

% version 3
% current methods: continuum model 1: bacteral density evolution; continuum model 2: two-phase field model; agent-based model 1: brownian dynamics + flow/director field; agent-based model 2: squirmer model + flow/director field;
%Active matter systems, exemplified by aqueous dispersions of self-propelled microswimmers, have garnered significant attention in recent years due to their ability to convert stored energy into mechanical motion~\cite{zhang2021}.

Suspensions of microswimmers are a protypical active matter, in which individual swimmers can self-propel by converting other forms of energy into mechanical work \cite{marchetti2013, saintillan2018, gompper2020}.
The individual and collective behaviors of microswimmers exhibit significant variability depending on several factors, including their species (which can be synthetic, such as robots or self-phoretic colloids, or living organisms, like bacteria and cytoskeletal extracts), population density, and the properties of the dispersing medium \cite{bechinger2016, chan2024}.
% Low density: single, pairwise interaction
% High density: active turbulence
Existing work has extensively investigated Newtonian suspensions of microswimmers. The scope of the continuous phase extends beyond isotropic solution such as water. In many biological and physiological contexts, microswimmers navigate through a complex, anisotropic medium. Notable examples include sperm cells in cervical mucus~\cite{suarez2006, kantsler2014}, bacteria or cells swimming in extracellular DNA biofilm matrix~\cite{smalyukh2008,flemming2010,lemon2017,repula2022}, synthetic microswimmers for drug delivery~\cite{gao2014, rao2015, wu2020}, and Trypanosoma brucei~\cite{masocha2004} moving through blood and lymphatic.
% extend the examples
The anisotropic and viscoelastic properties of the environment play a crucial role in modulating the dynamics of the dispersed swimmers, thereby offering a potential approach to control their dynamical behaviors~\cite{patteson2015,gomezsolano2016,tung2017,yuan2018,ishimoto2018,liu2021Viscoelastic}. 

A typical anisotropic medium is liquid crystal (LC). In the recent decade, non-toxic lyotropic chromonic LCs have been utilized to control the swimming behavior of dispersed bacteria~\cite{peng2016}, including \textit{E. coli}~\cite{kumar2013}, \textit{B. subtilis}~\cite{zhou2014,sokolov2015,peng2016,koizumi2020,genkin2017,genkin2018,turiv2020}, and \textit{P. mirabilis}~\cite{mushenheim2013,mushenheim2015,trivedi2015}. This type of composite system is named ``living nematic'' (LN)~\cite{zhou2014}. In the dilute limit, these bacteria tend to swim along the local nematic director due to a strong anisotropic viscous coefficient~\cite{zhou2014}. However, theoretical analyses and simulations indicate that the microswimmer's type, whether being a ``pusher'' or a ``puller''~\cite{lintuvuori2017,gautam2024}, and its surface anchoring type and strength~\cite{lintuvuori2017,chi2020}, can modify its preferred swimming direction. 
When the microswimmers are dense enough, they can interact with each other in an anisotropic manner~\cite{mushenheim2013,sokolov2015}. But more importantly, they can generate an active stress, which can induce elastic distortions to the background nematic, giving rise to the so-called ``active turbulent'' state, which is characterized by persistent proliferation, motion, and annihilation of topological defects~\cite{zhou2014, peng2016,genkin2017, genkin2018}. In a quasi two-dimensional (2D) film, extensile microswimmers can induce bend instability in an otherwise uniform nematic~\cite{zhou2014, peng2016, genkin2017, genkin2018,koizumi2020, turiv2020}. Whereas in a three-dimensional (3D) cell, they can trigger a twist-bend instability as a result of the competition between different elastic modes and the active stress~\cite{gautam2024}.

To control spontaneous flows and microswimmer dynamics in LNs, surface patterning technique has been proven promising~\cite{peng2016,genkin2017, genkin2018,koizumi2020}, which can also suppress or stabilize instability patterns~\cite{genkin2017,turiv2020}. 
Surface alignment (anchoring) condition can dictate the 3D trajectories of the microswimmers through controlling the bulk nematic field~\cite{mushenheim2015,zhou2017}. If the substrates of a nematic cell impose a planar anchoring condition, bacteria tend to swim within the midplane of the cell. Whereas if the substrates favor homeotropic anchoring, bacteria tend to swim in the bulk or form coordinated, trainlike groups~\cite{mushenheim2015,zhou2017}. Hybrid cells with planar and homeotropic surfaces further create nontrivial 3D trajectories, confining bacteria near the planar plate where splay deformation dominates~\cite{mushenheim2015}.

%Existing modeling efforts use either particle-based or continuum 

To elucidate bacterial dynamics in LNs, various types of continuum models have been developed. Spherical or ellipsoidal squirmer model has been used to represent individual microswimmers in an LN~\cite{lintuvuori2017,gautam2024,chi2020}. \textcolor{black}{By placing the sphere between two active force centers asymmetrically, the motions of microswimmers in the anisotropic fluid are well captured~\cite{daddi2018}}. 
When the bacterial size and shape are not essential, point-particle approximation can be adopted to study their collective dynamics~\cite{ryan2013,koizumi2020}. At a more coarse-grained level, an advection-diffusion equation for the bacterial concentration field can be used to describe the temporal behavior of bacterial distribution in an LN~\cite{genkin2017,genkin2018}. Alternatively, bacteria-rich regions can be modeled as an active nematic phase in a biphasic LC model~\cite{turiv2020}. 
These models provide useful insights into individual microswimmer behavior in an LC and their concentration modulation by topological defects. 

\textcolor{black}{In addition to continuum models, mesoscopic approaches that account for thermal fluctuations of nematic systems, such as multiparticle collision dynamics (MPCD) for liquid crystals ~\cite{lee2015stochastic,shendruk2015multi,mandal2019multiparticle,hijar2020dynamics,mandal2021} or active nematics ~\cite{kozhukhov2022,macias2023,kozhukhov2024}, have been recently developed and demonstrated to be effective. 
In particular, a fully particle-based method that integrates the nematic MPCD algorithm with a molecular dynamics (MD) scheme for spherical squirmers has been recently introduced~\cite{mandal2021,mandal2025}. This method has successfully reproduced the orientation dynamics of squirmers in nematic liquid crystals~\cite{mandal2021} and revealed the intricate dynamics of single and multiple squirmers in a homeotropically aligned nematic liquid crystal cell~\cite{mandal2025}.}
%In spite of these existing modeling efforts, it is still challenging to simulate a large number of generic microswimmers dispersed in a 3D nematic LC. 

Despite a relatively good understanding of LNs in dilute and dense limits, as revealed by the aforementioned models, and the promising applications of patterned surfaces to guide bacterial dynamics, it remains unclear what happens under certain conditions, e.g., in the semi-dilute condition, instability patterns can stop growing and become stabilized. Additionally, a related question remains unanswered: how do dilute-limit bacterial dynamics transition into dense-limit collective behavior as bacterial concentration gradually increases?

%Previous studies have established agent-based models by using the squirmer model to describe the perturbation of the surrounding flow by spherical~\cite{lintuvuori2017,gautam2024} or ellipsoidal microswimmers~\cite{chi2020}. Point particles can also be used to model microswimmers when their size and shape are not essential~\cite{ryan2013,koizumi2020}. Continuum models can be more suitable to study the collective behaviour of a large number of microswimmers, ignoring the individual responses and interactions, the advection-diffusion equation~\cite{genkin2017,genkin2018} are well validated. The phase-field model is also an option by regarding bacterial aggregation as a nematic phase with constant activity surrounded by an isotropic phase~\cite{koizumi2020}.

To address these open questions, here we develop a hybrid simulation method, which can quantitatively describe nematodynamics while accounting for arbitrary number of microswimmers. Specifically, we combine an agent-based method~\cite{ryan2013,koizumi2020} and a hydrodynamic model~\cite{denniston2001} to simulate self-propelled point-like particles in 2D and/or 3D nematic LCs. A similar hybrid approach has been successful in modeling aqueous solutions of microswimmers~\cite{li2019, li2024}.
We validate our computational model by reproducing key phenomena in previous experimental and numerical results~\cite{peng2016,koizumi2020,turiv2020,genkin2017}. We further use the model to examine the dynamics of microswimmers in a 3D nematic cell with and without topological defects. We also take into account the reversal of bacterial swimming direction in our simulation and find that depending on the reversal time scale, they are not necessarily accumulating in planar-anchoring walls. Taken together, we have developed a 3D simulation method to model LNs, facilitating further research in using anisotropic media to guide microswimmers.

%the control of the swimming bacteria in liquid crystals with spiral~\cite{peng2016,koizumi2020} and periodic splay-bend~\cite{turiv2020} patterned cell, and the accumulation and depletion around the cores of $\pm 1/2$ defects~\cite{genkin2017}. 
%This approach offers distinct advantages in simulating the onset of collective behaviors through gradual density increases, while also facilitating the study of activity-induced instability development. 
%Our model can be extended to 3D, allowing us to consider a nematic cell with patterns on both the top and bottom surfaces, and to study the coupling between the microswimmers and 3D disinclination lines. 

\section{Model}
\subsection{Agent-based Model to Simulate Microswimmers}
We model active particles (e.g., microswimmers in the experiment) as \textcolor{black}{active Brownian particles (ABPs)} with a constant self-propelled velocity $v_0$ along a direction represented by a unit vector $\hat{\mathbf{p}}_i$ for the $i$'th swimmer. In this work, we occasionally use swimmer and particle interchangeably to refer to microswimmer. Its equation of motion is governed by an over-damped active Brownian dynamics equation
\begin{equation}
\partial_t \mathbf{r}_i=v_0 \hat{\mathbf{p}}_i+\gamma_{L J} \mathbf{F}_i+\mathbf{u}, \label{equ:pos}
\end{equation}
where $\mathbf{u}$ is the velocity vector representing the background flow, and $\mathbf{F}_i=-\nabla_{\mathbf{r}_{i}} \sum_{i<j}V\left(\mathbf{r}_{ij}\right)$ is the Lennard--Jones-like force to describe the excluded volume interactions between particle $i$ and $j$, which is derived from the following potential function
\begin{equation}
V(\mathbf{r}_{ij})= \begin{cases}4 \varepsilon_{L J}\left[\left(\frac{\sigma}{r_{ij}}\right)^{12}-\left(\frac{\sigma}{r_{ij}}\right)^6\right], & r_{ij} \leq 2^{1 / 6} \sigma \\ 0, & r_{ij}>2^{1 / 6} \sigma\end{cases},
\end{equation}
where $r_{ij}=|\mathbf{r}_{ij}|$ is the distance between the two microswimmers, $\varepsilon_{L J}$ is the interaction energy scale, and $\sigma$ is the characteristic (excluded-volume) size of the microswimmer. 

\textcolor{black}{For ``pusher'' type microswimmers, they} tend to align with the local director field denoted by a unit vector $\hat{\mathbf{n}}$~\cite{lintuvuori2017,mandal2019multiparticle}. A torque is acted on the orientation vector $\hat{\mathbf{p}}_i$ about an axis $\mathbf{w}_s$, which can be expressed as $
\mathbf{w}_s = 2\gamma_B (\hat{\mathbf{p}}_i \cdot \hat{\mathbf{n}})(\hat{\mathbf{p}_i} \times \hat{\mathbf{n}}) = |\mathbf{w}_s| \hat{\mathbf{s}}$, where $\gamma_B$ is a coupling parameter, $\hat{\mathbf{s}}$ is a unit vector denoting the rotation axis, and the magnitude $|\mathbf{w}_s|$ is the angular velocity of the orientation vector $\hat{\mathbf{p}}_i$, namely $\Delta\theta = |\mathbf{w}_s| \Delta t$ for an infinitesimal duration of time $\Delta t$. To update $\hat{\mathbf{p}}_i$ in the simulation, we introduce the rotation matrix $\mathcal{R}$,
%defined by the vector $\mathbf{s}$. This vector can be expressed as $\mathbf{w}_s = 2\gamma_B (\hat{\mathbf{p}}_i \cdot \hat{\mathbf{n}})(\hat{\mathbf{p}_i} \times \hat{\mathbf{n}}) = |\mathbf{w}_s| \hat{\mathbf{s}}$, where $\gamma_B$ is a coupling parameter. The magnitude $|\mathbf{w}_s|$ defines the angular rotation rate, while the unit vector $\hat{\mathbf{s}}$ provides the axis of rotation. The rotation matrix $\mathcal{R}$ encodes the angular rotation by an angle $\Delta\theta = |\mathbf{w}_s| \Delta t$, 
which is implemented using the Rodrigues formalism~\cite{dai2015}:
\begin{equation}
\mathcal{R}(\Delta\theta) = e^{\Delta\theta \pmb{s}_{\times}} = \mathbf{I} + \sin\Delta\theta \pmb{s}_{\times} + (1 - \cos\Delta\theta)\pmb{s}_{\times}^2, \label{equ:RD}
\end{equation}
where the skew-symmetric matrix $[\pmb{s}_{\times}]$ is defined as:
\begin{equation}
[\pmb{s}_{\times}] = \begin{bmatrix}
0 & -s_z & s_y \\
s_z & 0 & -s_x \\
-s_y & s_x & 0
\end{bmatrix}.
\end{equation}
The updated orientation vector at time $t + \Delta t$ is then given by $\hat{\mathbf{p}}_i(t + \Delta t) = \mathcal{R}(\Delta\theta)\hat{\mathbf{p}}_i(t),$ and the time derivative of $\hat{\mathbf{p}}_i$ aligns with the angular velocity as $\partial_t \hat{\mathbf{p}}_i = \mathbf{w}_s \times \hat{\mathbf{p}}_i.$

Additionally, we consider the reorientation of the microswimmer in the presence of a background flow field. This reorientation is described by $
\partial_t \hat{\mathbf{p}}_i = \frac{1}{2} \pmb{\omega} \times \hat{\mathbf{p}}_i + B\left(\mathbf{E} \cdot \hat{\mathbf{p}}_i - (\hat{\mathbf{p}}_i \cdot \mathbf{E} \cdot \hat{\mathbf{p}}_i)\hat{\mathbf{p}}_i\right),
$
which originates from Jeffrey's equation~\cite{ryan2013,ishimoto2023}. Here, $\pmb{\omega} = \nabla \times \mathbf{u}$ represents the vorticity, and $\mathbf{E} = \frac{1}{2}(\nabla \mathbf{u} + (\nabla \mathbf{u})^T)$ is the strain-rate tensor. The parameter $B$, known as the Bretherton parameter, is defined as $B = (e^2 - 1)/(e^2 + 1)$, where $e$ is the aspect ratio of the microswimmers.

Finally, the evolution of the orientation vector $\hat{\mathbf{p}}_i$ for particle $i$ is expressed as:
\begin{equation}
\begin{aligned}
\partial_t \hat{\mathbf{p}}_i = & (\mathbf{w}_s + \frac{1}{2} \pmb{\omega}) \times \hat{\mathbf{p}}_i + B\left(\mathbf{E} \cdot \hat{\mathbf{p}}_i - (\hat{\mathbf{p}}_i \cdot \mathbf{E} \cdot \hat{\mathbf{p}}_i)\hat{\mathbf{p}}_i\right)\\
&+ D_r \xi(t) \hat{\mathbf{p}}_i^{\perp}, \label{equ:3D_Ori}
\end{aligned}
\end{equation}
where the third term introduces a white noise, $\xi(t)$ with time de-correlated and normalized amplitude $\langle\xi(t)\xi(t') \rangle=\delta(t,t')$, and $\hat{\mathbf{p}}_i^{\perp}$ represents a unit vector perpendicular to $\hat{\mathbf{p}}_i$.

In 2D simulations, the orientation vector for particle $i$ can be represented by its orientation angle $\theta_i^p$ via $\hat{\mathbf{p}}_i=\left(\cos\theta_i^p,\sin\theta_i^p\right)$. Likewise, the nematic director can also be rewritten in terms of the director angle $\theta$ through $\hat{\mathbf{n}}=\left(\cos\theta,\sin\theta\right)$. Thus, the orientational equation of motion Eq.~\ref{equ:3D_Ori} can be simplified as
%By using $p_x \partial_t p_y-p_y \partial_t p_x=\partial_t \theta_p$, we can map the 3D expression to a 2D formula. In 2D, the bacterial orientation can be written as $\hat{\mathbf{p}}=\left(\cos\theta_p,\sin\theta_p\right)$, and the nematic director becomes $\hat{\mathbf{n}}=\left(\cos\theta,\sin\theta\right)$. The orientational equation can thus be simplified as
\begin{equation}
\begin{aligned}
\partial_t \theta_i^p=&\gamma_B\sin \left[ 2(\theta-\theta_i^p)\right]+\frac{1}{2}{\omega}_z\\&+\frac{1}{2}B\left[\left(E_{yy}-E_{xx}\right)\sin(2\theta_i^p)+2E_{xy}\cos(2\theta_i^p)\right]\\
&+D_r \xi(t). \label{equ:2D_Ori}
\end{aligned}
\end{equation}

\subsection{Continuum Model to Solve the Nematic Directors and the Velocity Field}
\label{sec:model}
We employ a hybrid lattice Boltzmann method (LBM) to perform simulations to solve the nematic director field and the flow field~\cite{zhang2016}. 
The microstructure of the nematic liquid crystal is characterized by a traceless, symmetric tensorial order parameter $\mathbf{Q}$. For a uniaxial nematic LC, $\mathbf{Q}=S(\mathbf{n} \mathbf{n}-\mathbf{I} / 3)$, where $S$ is the scalar order parameter.
The governing equation of the $\mathbf{Q}$-tensor is described by the Beris--Edwards equation~\cite{beris1994,denniston2001}:
\begin{equation}
\partial_t \mathbf{Q}+\mathbf{u} \cdot \nabla \mathbf{Q}-\mathbf{S}=\Gamma \mathbf{H}. \label{B-S Equ}
\end{equation}
By defining the symmetric part and the antisymmetric part of the velocity gradient tensor $\mathbf{E}=\frac{1}{2}\left(\nabla \mathbf{u}+(\nabla \mathbf{u})^T\right)$ and the vorticity tensor $\mathbf{W}=\frac{1}{2}\left(\nabla \mathbf{u}-(\nabla \mathbf{u})^T\right)$, the corotation term that determines the alignment of the elongated units in response to the velocity gradient is given by $\mathbf{S}=(\xi \mathbf{E}+\mathbf{W}) \cdot\left(\mathbf{Q}+\frac{\mathbf{I}}{3}\right)+\left(\mathbf{Q}+\frac{\mathbf{I}}{3}\right) \cdot(\xi \mathbf{E}-\mathbf{W})-2 \xi\left(\mathbf{Q}+\frac{\mathbf{I}}{3}\right)(\mathbf{Q}: \boldsymbol{\nabla} \mathbf{u})$. $\Gamma$ controls the relaxation rate of the director;  $\mathbf{H}$ denotes the molecular field which drives the system towards the thermodynamic equilibrium,  ${\mathbf{H}}=-\left(\frac{\delta F}{\delta\mathbf{Q}}-\frac{\mathbf{I} }{3}\mathrm{Tr}\left(\frac{\delta F}{\delta\mathbf{Q}}\right)\right)$, where the total free energy $F=\int_V\left(f_{\text {LdG }}+f_{\text {ela }}\right) \mathrm{d} V+\int_{\partial V} f_s \mathrm{d} S$ contains the Landau--de Gennes free energy density
\begin{equation}
\begin{aligned}
f_{\mathrm{LdG}}= &\frac{A_0}{2}\left(1-\frac{U}{3}\right) \operatorname{Tr}\left(\mathbf{Q}^2\right)-\frac{A_0 U}{3} \operatorname{Tr}\left(\mathbf{Q}^3\right) \\ &+\frac{A_0 U}{4}\left(\operatorname{Tr}\left(\mathbf{Q}^2\right)\right)^2,
\end{aligned}
\end{equation}
the elastic energy density
\begin{equation}
f_{\text {ela }}=\frac{1}{2} L Q_{i j, k} Q_{i j, k},
\end{equation}
and the surface anchoring energy $f_s=\frac{1}{2} W_s\left(\mathbf{Q}-\mathbf{Q}^s\right)^2$, with $\mathbf{Q}^s$ denoting the preferred order parameter by the surface. 
{In our simulations, all length scales are normalized by the nematic coherence length $\xi_N = \sqrt{L / A_0}$. The unit of time is $\tau = \gamma \xi_N^2 / L$ (where $\gamma$ is the rotational viscosity), and the unit of energy is $E_0 = L \xi_N$. }%Anchoring strength is expressed in reduced units as $\hat{W} = \xi_N / \xi_W$.}%In the simulation, we can normalize all length scale as the coherence length $\xi_N = \sqrt{\frac{L}{A_0}}$. and extrapolation length $\xi_W = {L / W_s}$.

The flow field obeys the incompressibility equation
\begin{equation}
\nabla \cdot \mathbf{u}=0,
\end{equation}
and the Navier--Stokes equation
\begin{equation}
\rho_f\left(\partial_t \mathbf{u}+\mathbf{u} \cdot \nabla \mathbf{u}\right)=\nabla \cdot\left(\pmb{\Pi}^\text{p}+\pmb{\Pi}^\text{a}\right).
\end{equation}
The passive stress tensor is defined as $\Pi_{\alpha \beta}^\text{p}= 2 \eta E_{\alpha \beta} -P_0 \delta_{\alpha \beta}-\xi H_{\alpha \gamma}\left(Q_{\gamma \beta}+\frac{1}{3} \delta_{\gamma \beta}\right)-\xi\left(Q_{\alpha \gamma}+\frac{1}{3} \delta_{\gamma \beta}\right) H_{\gamma \beta}+2 \xi\left(Q_{\alpha \beta}+\frac{1}{3} \delta_{\alpha \beta}\right) Q_{\gamma \epsilon} H_{\gamma \epsilon} -\partial_\beta Q_{\gamma \epsilon} \frac{\delta F}{\delta \partial_\alpha Q_{\alpha \epsilon}}+Q_{\alpha \gamma} H_{\gamma \beta}-H_{\alpha \gamma} Q_{\gamma \beta}-\zeta Q_{\alpha \beta},$
where $\eta$ is the isotropic viscosity, $\rho_f$ is the fluid density, $\xi$ is the flow-aligning parameter, and $P_0$ denotes the hydrostatic pressure. 
The active stress tensor stems from the microswimmers, $\pmb{\Pi}^\text{a}=-\zeta_B \sum _if(\mathbf{r} - \mathbf{r}_i)\mathbf{Q}^{\mathrm{B}}_i $, where $\mathbf{Q}_i^{\mathrm{B}}=\left(\hat{\mathbf{p}}_i \hat{\mathbf{p}}_i-{\mathbf{I}}/{3}\right)$ denotes the orientation of the microswimmers. $\alpha_B(\mathbf{r})=\zeta_B \sum_i f(\mathbf{r} - \mathbf{r}_i)$ represents the spatially varying activity determined by the distributions of the microswimmers. Here, the kernel $f(\mathbf{r} - \mathbf{r}_i)$ describes the spatial influence of a microswimmer at position $\mathbf{r}_i$ on the activity at position $\mathbf{r}$. 

\begin{figure*}[htp]
	\centering
	%\vspace{\spaceAboveFigure}
	\includegraphics[width=\linewidth]{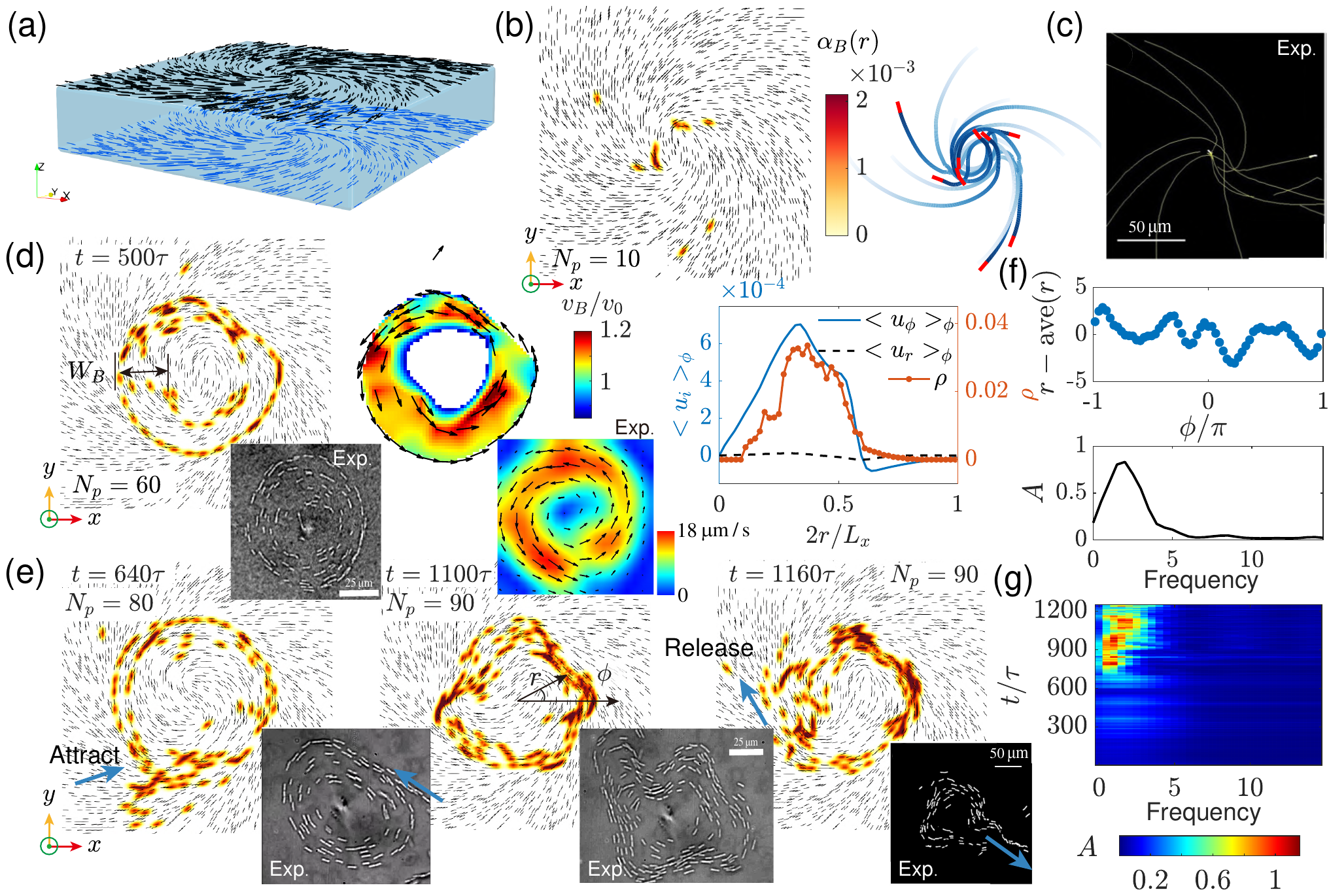}
	\vspace{\spaceBelowFigure}
	\phantomsubfloat{fig:spiral:a}
    \phantomsubfloat{fig:spiral:b}
    \phantomsubfloat{fig:spiral:c}
    \phantomsubfloat{fig:spiral:d}
    \phantomsubfloat{fig:spiral:e}
    \phantomsubfloat{fig:spiral:f}
    \phantomsubfloat{fig:spiral:g}
    %\vspace{-2\baselineskip}% Remove extra line inserted by subfloat
    \vspace{-1.5\baselineskip}
    \caption{Microswimmer dynamics on a spiral $+1$ defect pattern. (a) The schematic of the simulation box with a spiral pattern on both top and bottom surfaces \textcolor{black}{with $L_x=95$, $L_y=95$, $L_z=20$, $\zeta_B=0.03$, and $l=1$}.
	(b) Configuration with small amount of bacteria $N_p=10$. Black sticks represent the nematic director, while color indicates the function $\alpha_B(\mathbf{r})$, signifying the positions of individual microswimmers. The right panel illustrates swimmer trajectories with red rods marking their final positions.
    (c) Experimental optical microscopy showing bacteria trajectories that follow the surface-imposed nematic directors at low concentration~\cite{koizumi2020}.
    (d) Numerical results of stable swirling at higher population $N_p=60$. The left and middle panel show the position and the particle velocity $v_B$ at $t=500\tau$. 
    \textcolor{black}{Insets: experimental observations of the stable swirling~\cite{peng2016}. The color map depicts bacterial velocity.}
    The right panel is the azimuthal average of the average flow field from $t=450 \tau$ to $t=600 \tau$ ($\tau = 100$ time steps in simulation) in tangential ($u_\phi$) and radial direction ($u_r$) as a function of the distance to the origin. There are $40$ swimmers initially, we add $20$, $20$, $10$ particles at $t=$ $300\tau$, $600\tau$ and $900\tau$, respectively. 
    (e) Numerical results initiated from $t=0$ of (d) The attraction of outer bacteria to the swirl at $t=640\tau$, undulation at $t=1100\tau$, and release from protrusions at $t=1160\tau$. The experimental snapshots are from Refs.~\cite{peng2016,koizumi2020}.
    \textcolor{black}{(f) Variation of the swirl’s radial distance from the center, $r$, as a function of the azimuthal angle $\phi$, as defined in (e). Bottom: The amplitude $A$ as a function of frequency, obtained via Fast Fourier transform (FFT). The data is extracted at $t = 1100\tau$.}
    \textcolor{black}{(g) The time-resolved frequency spectrum illustrating the transition from stable swirling to undulation.}
    }
    \vspace{\spaceBelowCaption}
    \label{fig:spiral}
\end{figure*}
These microswimmers are characterized as pushers ($\zeta_B>0$)~\cite{darnton2004}, generating hydrodynamic \textcolor{black}{point} force dipoles typical of many flagellated bacteria in isotropic solutions~\cite{edwards2009,drescher2010,hintsche2017,singh2017,park2017,alapan2018}. 
The axisymmetric shape of the microswimmer is described by an anisotropic-Gaussian-like distribution ~\cite{koizumi2020}
\begin{equation}
\begin{aligned}
f(\mathbf{r} - \mathbf{r}_i) =& \frac{1}{(2 \pi)^{d / 2} g_{\|} g_{\perp}^{d-1}} \exp\left[ -\frac{\left((\mathbf{r} - \mathbf{r}_i) \cdot \hat{\mathbf{p}}_i\right)^2}{g_{\|}^2} \right. \\
&\left. - \frac{\left|(\mathbf{r} - \mathbf{r}_i) - \left((\mathbf{r} - \mathbf{r}_i) \cdot \hat{\mathbf{p}}_i\right) \hat{\mathbf{p}}_i\right|^2}{g_{\perp}^2} \right],
\end{aligned}
\end{equation}
where $g_{||}$ and $g_{\perp}$ ($g_{||}/ g_{\perp} = e$, $g_{\perp} = l$) define the width of the distribution along the rotational symmetry axis and the direction perpendicular to it, respectively. 

In the following simulations, parameters used are: $\gamma_B = 0.004$, $\varepsilon_{LJ}=4.167\times10^{-4}$, $\gamma_{LJ}=0.1$, $D_r=1\times10^{-5}$, {$e = 2.2$}, $\Gamma=0.1$, $A_0 = 0.01$, $U = 0.35$, $L=0.01$, $W_s=1$, $\rho_f=1$, $\eta=0.33$, and {$\xi=0.3$}.

\section{Results and Discussion}

%\subsection{Benchmarks}
\subsection{Unidirectional Swimming using Patterned Surfaces}
\label{sec:Pattern}
\subsubsection{Circulation and Undulation in Spiral-patterned Cell}
We first study the dynamics of microswimmers over a spiral-patterned surface in a thin nematic cell. 
\textcolor{black}{Previous literature~\cite{peng2016,koizumi2020} show that in dilute dispersions, bacteria follow nonpolar spiral trajectories along the pre-imposed molecular orientation; but as their concentration increases, they form unipolar circular swirls which can further grow into an undulating shape at higher densities. In the following section, we use our hybrid method to reproduce the states transitions and quantify the transition from stable swirling to undulation state with spectrum calculations. We also pay close attention to details such as the flow field and the dynamic formation and breakdown of the swirl. These aspects were absent in previous simulation analyses~\cite{koizumi2020}.} 
The simulation setup is shown in \cref{fig:spiral:a}, where we apply the same clockwise-spiral $+1$ defect pattern on both surfaces with strong anchoring ($W=1$). For simplicity, the dynamics of the microswimmers are governed by the 2D equations \ref{equ:pos} and \ref{equ:2D_Ori}.

%In \cref{fig:spiral:b}, a small population of microswimmers (population number is $N_p=10$)
When the microswimmer population number $N_p$ is low (up to $N_p=10$), they are found to roughly align with the designated nematic field, and can move away from the spiral center after entering it (\cref{fig:spiral:b}). This behavior is consistent with the experiment of a dilute LN~\cite{koizumi2020} (\cref{fig:spiral:c}). 
As the population increases to $N_p=60$, the microswimmers do not faithfully follow the clockwise spiral pattern; instead, they are persistently swirling counterclockwise, generating a stable vortex flow (\cref{fig:spiral:d}).
The central $+1$ defect splits into two $+1/2$ defects that orbit the center counterclockwise due to the active flow. The colormap of the swimmer velocity matches well with the experiment~\cite{peng2016} (\cref{fig:spiral:e}). 
We further measure the azimuthally averaged flow velocity in the polar coordinate, $\langle u_r\rangle_{\phi}$ and $\langle u_\phi\rangle_{\phi}$ (\cref{fig:spiral:d}, right panel). The azimuthal velocity peak matches with the density distribution of the swimmers, which also exhibits a finite width (\cref{fig:spiral:d,fig:spiral:e}). 
\textcolor{black}{This occurs because the local active stress responsible for generating spontaneous flow is proportional to the swimmer density; thus, the region with the highest swimmer density is expected to exhibit the highest flow velocity. 
Our method captures this phenomenon, which is absent in previous particle-based models where swimmers are confined to a single-particle-width circular ring with uniform swimming speed}~\cite{koizumi2020}.
%plot the average flow field in the azimuthal direction ($u_\phi$) and radial direction ($u_r$). While $u_r$ remains nearly zero, $u_\phi$ peaks within the swirl and decreases sharply near its outer boundary. The non-zero $u_\phi$ near the center drives the orbiting motion of the two $+1/2$ defects. 

\begin{figure*}[htp]
	\centering
	%\vspace{\spaceAboveFigure}
	\includegraphics[width=\linewidth]{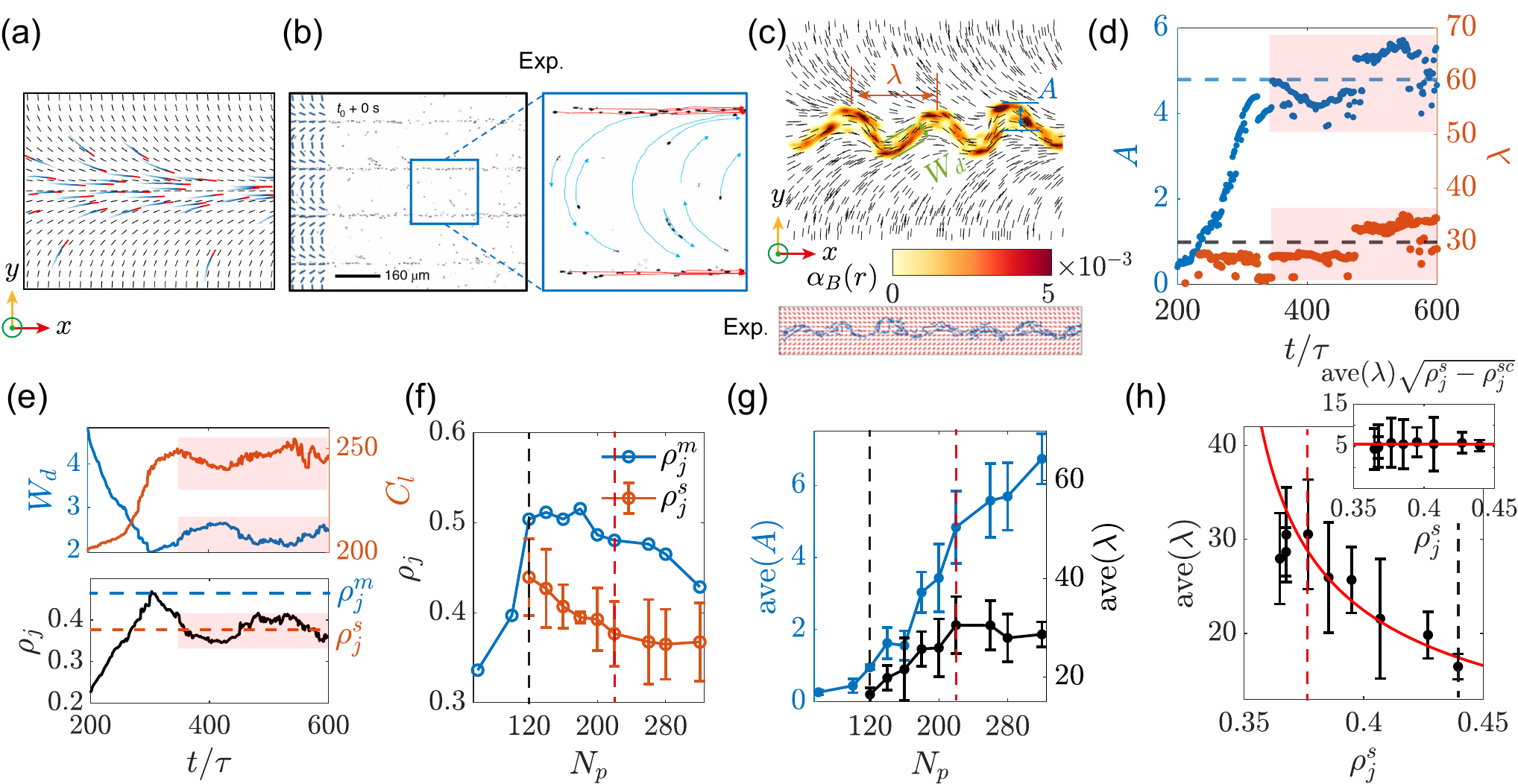}
	\vspace{\spaceBelowFigure}
	\phantomsubfloat{fig:PolarJet:a}
    \phantomsubfloat{fig:PolarJet:b}
    \phantomsubfloat{fig:PolarJet:c}
    \phantomsubfloat{fig:PolarJet:d}
    \phantomsubfloat{fig:PolarJet:e}
    \phantomsubfloat{fig:PolarJet:f}
    \phantomsubfloat{fig:PolarJet:g}
    \phantomsubfloat{fig:PolarJet:h}
    %\vspace{-2\baselineskip}% Remove extra line inserted by subfloat
    \vspace{-2\baselineskip}
    \caption{Polar jets and their undulation over a periodic splay-bend ``C''-pattern \textcolor{black}{with $L_x=200$, $L_y=75$, $L_z=12$, $\zeta_B=0.03$, and $l=1$}.
	    (a) Simulated microswimmer trajectories at low concentration with $N_p=60$. Bacteria concentrate in the splay regions and occasionally cross the bend region.
        (b) Experimental microscope images of bacterial trajectories~\cite{turiv2020} confirming our findings in (a).
        (c) A fully-developed stabilized undulation of the jet. Top: simulation result for $N_p=220$ at $t=500\tau$. Graphical definitions of the undulation wavelength $\lambda$, undulation amplitude $A$, and jet width $W_d$ are given.
        Bottom: Experimental results from Ref.~\cite{turiv2020}. The red sticks represent the predesigned surface nematic field and blue sticks represent the bacteria. 
        (d) Time evolution of $A$ and $\lambda$ of the jet in (c). 
        (e) Time evolution of $W_d$, $C_l$ and $\rho_j$ of the jet in (c). 
        (f) The maximum value of swimmer density in the jet $\rho_j^m$ and the average density in the stable jet $\rho_j^s$ versus $N_p$.
        (g) $A$ and $\lambda$ of the wave in the fully developed region as functions of $N_p$. 
        (h) $\lambda$ as a function of $\rho_j^s$. The red solid line is the fit $\lambda \propto 1/\sqrt{\rho_j^s-\rho_j^{sc}}$. Inset: The collapse of the data $\textnormal{ave}(\lambda) /\sqrt{\rho_j^s-\rho_j^{sc}}$.  }
    \vspace{\spaceBelowCaption}
    \label{fig:PolarJet}
\end{figure*}
As demonstrated in \cref{fig:spiral:f} and {Supplementary Video 1}, once a stable swirl of $N_p=80$ microswimmers is established at $t=640\tau$, 10 more swimmers are added to the simulation box. Then the swirl of $N_p=90$ particles begins to undulate due to the bend instability associated with the extensile activity~\cite{zhou2014,genkin2017,peng2016,koizumi2020}. 
%newly introduced bacteria are attracted and assimilated into the swirl~\cite{peng2016,koizumi2020}. The swirl begins to undulate with further bacterial increase around $t=1100\tau$ in our simulation, due to the bend instability associated with the extensile activity~\cite{zhou2014,genkin2017,peng2016,koizumi2020}. 
Some swimmers evade the most prominent protrusions, while the others continue to swirl ($t=1160\tau$ in \cref{fig:spiral:f})---this again reproduces the experiment~\cite{koizumi2020}. 
\textcolor{black}{We calculate the spectrum of the swirl to quantitatively distinguish between the stable swirling state and the undulation state in \cref{fig:spiral:f,fig:spiral:g}. For stable swirling, the band is circular. However, instability induces undulations, whose dominant frequency can be determined using the Fast Fourier Transform (FFT), as shown in~\cref{fig:spiral:f}. The dominant wavelength at $t=1100\tau$ is given by: $\lambda=\frac{2\mathrm{ave}(r)}{\mathrm{dominant \ frequency}}\approx23$. We plot the time-resolved spectrum in~\cref{fig:spiral:g}to show the states transition, where the periodic wave gradually appears around $t=650 \tau$ ($N_p=80$), and the dominant frequency remains steady around $2$, until the wave structure is destroyed by the release of the swimmers.}
%Compared with the advection-diffusion model developed by Genkin \textit{et al.}~\cite{genkin2017} and applied by Koizumi \textit{et al.}~\cite{koizumi2020}, our model performs better in the details. For example, our model captures the finite width of the circulation swirl and the swimming velocity variation along the radial direction; we faithfully reproduce how the several bacteria being attracted or released from the swirl, which is obscure in the continuum model; 
%We obtain the stable unidirectional circulation and undulation when the angle between the nematic background and the radial direction $\phi_0=45 \degree$, which is trickier to reach than the larger value of $\phi_0$ ($\phi_0$ falls in $[65 \degree, 85 \degree]$ in the Ref.~\cite{koizumi2020}) where the enlarged bend configuration helps to trap the bacteria in circle.

\subsubsection{Polar Jets and Undulation in Periodic C-patterned Cell}
We next examine a different system where a periodic C-patterned surface is used to guide a unidirectional bacterial flow (namely a ``bacterial jet'') in an LN. As reported in Ref.~\cite{turiv2020}, 
\textcolor{black}{bacteria are expelled from the bend regions and condense into stable, unidirectional polar jets along the splay regions, until bend instability emerges when the bacterial concentration exceeds a threshold. Our analysis offers a quantitative examination of the development and stabilization of the wave structure, along with a fitting to demonstrate the dependence of the characteristic wavelength on swimmer density. These aspects are absent in previous simulation analyses~\cite{turiv2020}.} 
When a small population of microswimmers ($N_p\le 100$) are over a periodic splay-bend pattern, they will migrate into the splay regions and form a unidirectional jet along the converging direction (i.e., the $+x$ direction, see \cref{fig:PolarJet:a}). 
Once the polar jet is established in the splay regions, the swimmers become trapped and rarely enter the bend region. 
These key features of the experiment~\cite{turiv2020} are well reproduced (\cref{fig:PolarJet:a,fig:PolarJet:b}). 
As more bacteria accumulate in the splay band ($N_p \ge 120$), the initially straight jet develops a periodic undulation, leading to the distortion of the nearby nematic field (\cref{fig:PolarJet:c} and Supplementary Video 2). This behavior is also clearly observed in the experiment~\cite{turiv2020}.
To quantify this transition, we introduce microswimmer concentration (or population density) within the jet defined as $\rho_j = N_p / (C_l W_d)$, where $C_l$ and $W_d$ are the contour length and the width of the jet, respectively (\cref{fig:PolarJet:c}).  
In \cref{fig:PolarJet:d,fig:PolarJet:e}, we plot the evolution of the amplitude $A$, the wavelength $\lambda$, $C_l$, $W_d$, and $\rho_j$ for $N_p = 220$. The swimmers gradually accumulate in the splay region and form a jet. At $t=200\tau$, $A$ and $C_l$ starts to grow, implying the onset of instability. In the wavy jet state, $A$ and $C_l$ of the undulation increase linearly before reaching a plateau, while its wavelength $\lambda$ remains stable (\cref{fig:PolarJet:d,fig:PolarJet:e}). As the undulation grows, the microswimmer density $\rho_j$ dilates, leading to a lower effective active stress. Therefore, the undulation will stop growing when the active stress is balanced by the associated free energy cost.
This wavy jet propagates at a constant phase velocity along the $+x$ axis once its structure is stabilized (Supplementary Video 2). 
The forward advancing of the undulation wave agrees with the experiment~\cite{turiv2020,Wu2025}, with details further discussed in a separate work~\cite{Wu2025}.

We record the maximum value of density in the jet as $\rho_j^m$ (when $W_d$ becomes minimum), and the average density of the steady wave $\rho_j^s$ as a function of $N_p$ in \cref{fig:PolarJet:f}. 
There is a competition between the bend-induced active torque and the free energy cost of the undulation. The higher the microswimmer concentration $\rho_j$ is, the higher the effective active stress would be. When $\rho_j^m$ reaches the threshold $\rho_j^0 \approx 0.50$, the bend undulation becomes unstable and will grow. 
The critical concentration for the onset of the undulation is approximately $\rho_j^0 \approx 0.50$, at which $N_p \approx 120$, corresponding to $4.45 \times 10^{10} \mathrm{m^{-2}}$ in SI units, which is close to the critical density reported in Ref.~\cite{turiv2020}, where $\rho_j^0 = 2.36 \times 10^{10} \mathrm{m^{-2}}$. 
Note that the threshold density for the spiral-pattern (\cref{fig:spiral}) is $\rho_j^0 \approx 0.1$, which is lower than that in the C-pattern system. This trend is consistent with the experiment: $\rho_j^0 \approx 0.57 \times 10^{10} \mathrm{m^{-2}}$ in the spiral-pattern system compared to $\rho_j^0 = 2.36 \times 10^{10} \mathrm{m^{-2}}$ in the C-pattern system~\cite{koizumi2020}. 
Both $\rho_j^s$ and $\rho_j^m$ decrease as $N_p$ increases once $\rho_j^m$ is above the threshold, as a result of fast growth of $C_l$ and $W_d$~\cite{turiv2020}. 

%The distorted nematic directors follow the wave-shaped jet of the swimmers, showing a larger angular deviation from the designed directions after the wave peak and a smaller deviation ahead of it.  This uneven rotation of the background nematic field facilitates the forward motion along $+x$ axis of the wave peaks. 

We further measure the steady values of $A$ and $\lambda$ for stable propagating jets, straight or wavy, and present their dependence on the microswimmer population $N_p$ and its concentration within the stable jet $\rho_j^s$ in \cref{fig:PolarJet:g,fig:PolarJet:h}. 
%Both $C_l$ and $W_d$ increase with $N_p$, resulting in a decrease in $\rho_j^m$ \cref{fig:PolarJet:f}~\cite{turiv2020}. 
When $120 \leq N_p \leq 220$, which corresponds to $0.37 \le \rho_j^s \le 0.44 $, the wavelength scales as $\lambda \propto 1/\sqrt{\rho_j^s - \rho_c^{sc}}$, with $\rho_j^{sc} = 0.34$. This relationship arises from the balance between active torque and elastic torque, consistent with Refs.~\cite{ramaswamy2010, zhou2014, turiv2020}. 
When $N_p > 220$, the concentration $\rho_j^s$ reaches a plateau around $0.36$, at which $W_d$ and $A$ increase slowly, while $\lambda$ is constantly around $29.05$ (\cref{fig:PolarJet:f,fig:PolarJet:h}). This suggests that the effective active stress is saturated when $N_p > 220$. 
A further increase in $N_p$ (further decrease in $\rho_j$) causes the swimmers to escape from the protrusions of the wave, leading to the rupture of the wave, as shown in Supplementary Video 3 for $\rho_j^s = 0.36$ ($N_p = 280$). 
We record values only before wave rupture, where $\lambda$ is determined by $\rho_j^s$.

Compared to the mean-field-like approach used in Ref.~\cite{turiv2020}, our particle-based method is numerically stable and can handle sharp boundaries between bacteria-rich phase (region) and bacteria-depleted phase.
%Compared to the advection-diffusion description of the microswimmers used in Ref.~\cite{turiv2020}, our approach provides an better solution for mapping the orientation of microswimmers to nematic directors. By considering nematic symmetry, our method avoids discontinuities at $\theta = \pm \pi/2$ (discussed in \cite{genkin2017}) and relaxes the restrictions associated with the nematic vector field representation~\cite{turiv2020}. Regarding the biphase-field framework described in Ref.~\cite{turiv2020}, our model addresses the challenges of describing highly dense microswimmer clusters with well-defined interfaces, where determining interfacial tension and the orientation of swimmers at the interface is nontrivial but crucial for capturing wave structures. 
The key advantage of our agent-based model lies in its ability to resolve individual swimmers. This allows us to faithfully capture the finite width of the undulation band and the rupture of the undulation, providing additional insights into the dynamics of the LN.

%Compared with Ref.~\cite{koizumi2020}, we replicate the previous experimental results, including the finite width of the bacterial swirls, and analyze the bacterial speed distribution, density profiles, and flow fields within the swirl. More importantly, our simulations capture the stabilized undulation by coupling the dynamics of microswimmers with the nematic field and active flow, leading to bend instability. At the resolution of individual swimmers, we faithfully reproduce behaviors such as attraction and escape, highlighting the critical role of the flow field in these interactions.

\subsection{Interactions with Topological Defects}

Defects widely exist in active turbulence when activity dominates over LC elasticity. 
In a 2D LN, bacteria tend to accumulate at the cores of $+1/2$ defects and are depleted from the cores of $-1/2$ defects, as demonstrated through both experimental observations and continuum simulations~\cite{genkin2017}. 
In 3D active nematics, disclination lines can form closed loops~\cite{duclos2020}. These loops are influenced by a self-propulsion velocity based on their local profiles, causing the defect loops to expand, contract, or buckle over time~\cite{binysh2020,long2021, kralj2023}. 
3D LNs are different from 3D active nematics, in which activity coefficient can be assumed uniform. Microswimmers in 3D LNs can aggregate at or deplete from certain defects, which can lead to modulated activity. Therefore, we expect richer defect dynamics in 3D LNs.

%Different from the previous study using a uniform activity, the aggregation and depletion of the bacteria result in a density modulation around the disinclination, leading to different behaviors of the disinclination lines. 
\begin{figure}[htp]
	\centering
	%\vspace{\spaceAboveFigure}
	\includegraphics[width=\linewidth]{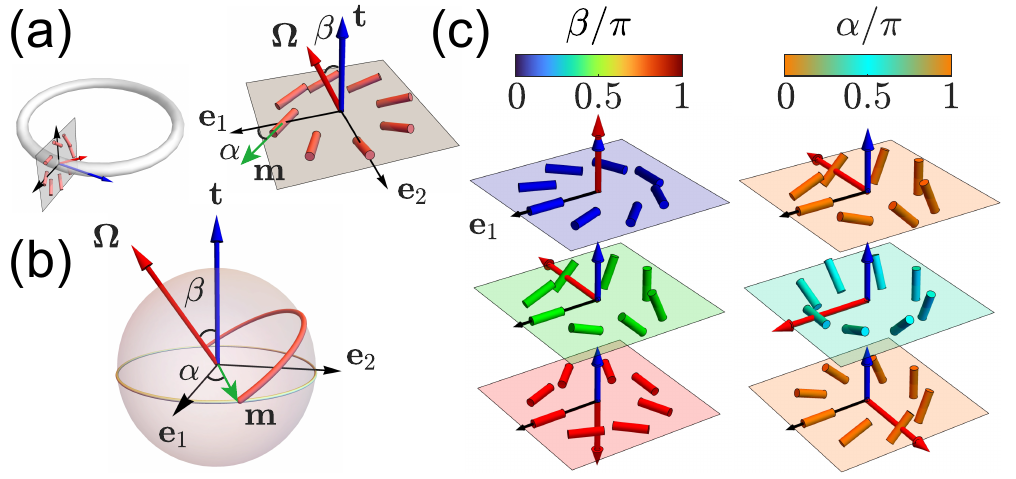}
	\vspace{\spaceBelowFigure}
	\phantomsubfloat{fig:order_param:a}
    \phantomsubfloat{fig:order_param:b}
    \phantomsubfloat{fig:order_param:c}
    %\vspace{-2\baselineskip}% Remove extra line inserted by subfloat
    \vspace{-1.5\baselineskip}
    \caption{Schematic representation of the local director profile for a winding of a line or loop defect. 
    \textcolor{black}{(a) A local director profile of a slice from a defect loop. }
    (b) Order parameter space with the pink path indicating the trajectory of the director ends starting from the origin. 
    (c) \textcolor{black}{Left: A $+1/2$ wedge defect transitions into a twist defect and then into a $-1/2$ wedge through the increase of $\beta$, with $\alpha=0$.
    Right: A pure-twist defect configuration with $\beta=\pi/2$ transforms from radial twist winding ($\alpha = 0$) to tangential twist winding ($\alpha = \pi/2$) and back to radial twist winding ($\alpha = \pi$).}
    }
    \vspace{\spaceBelowCaption}
    \label{fig:order_param}
\end{figure}

\begin{figure*}[htp]
	\centering
	%\vspace{\spaceAboveFigure}
	\includegraphics[width=0.76\linewidth]{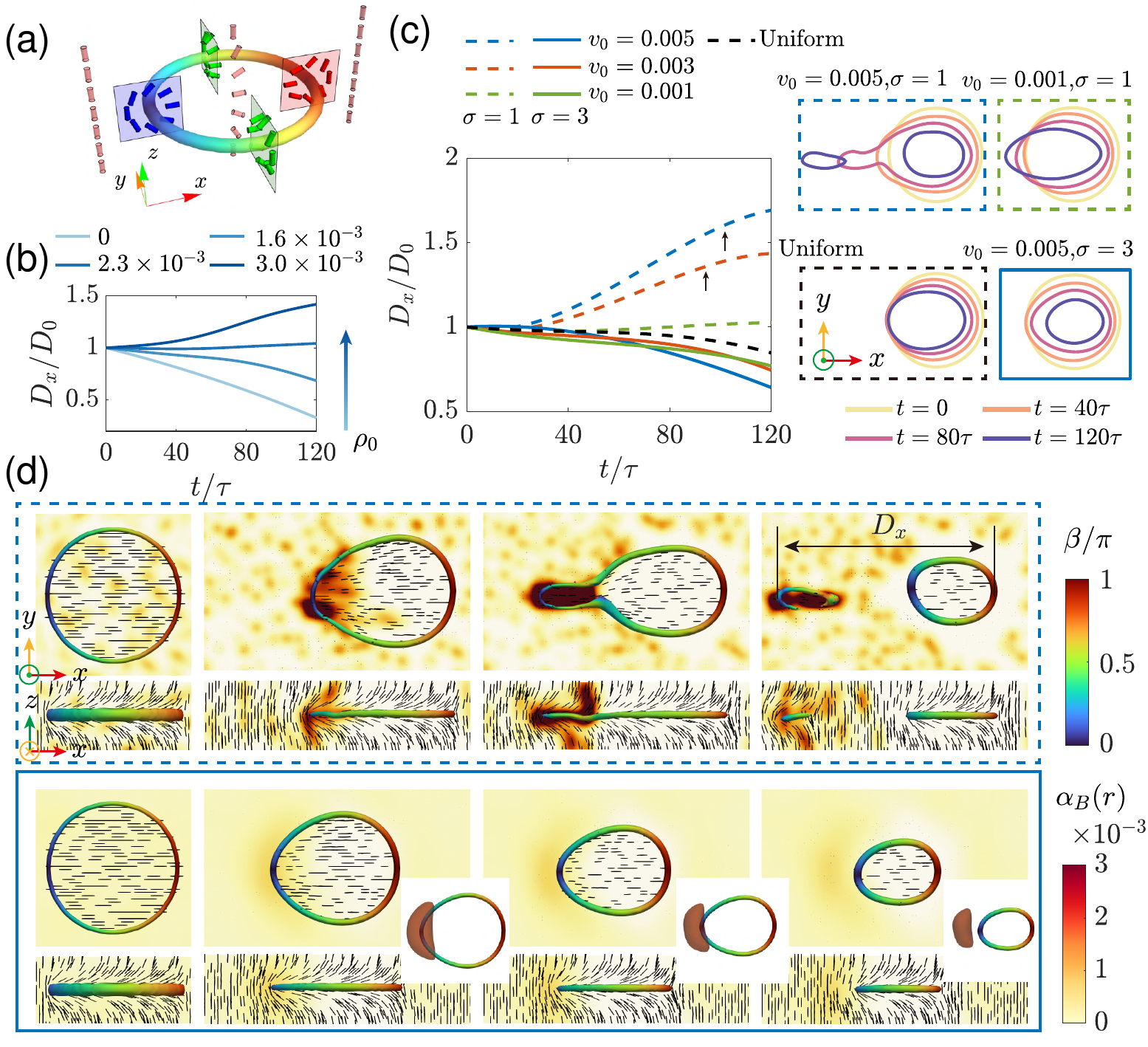}
	\vspace{\spaceBelowFigure}
	\phantomsubfloat{fig:LoopS:a}
    \phantomsubfloat{fig:LoopS:b}
    \phantomsubfloat{fig:LoopS:c}
    \phantomsubfloat{fig:LoopS:d}
    %\vspace{-2\baselineskip}% Remove extra line inserted by subfloat
    \vspace{-0.5\baselineskip}
\caption{Dynamics of a wedge-twist loop defect in a homeotropic-anchoring cell \textcolor{black}{with $L_x=160$, $L_y=80$, $L_z=22$, $\zeta_B=0.3$, and $l/\sigma=2$}. (a) The schematic of the loop defect. The color of the loop indicates the angle $\beta$ (colorbar is shown in panel (c)). \textcolor{black}{(b) The evolution of the lateral size $D_x$ (defined in (d)) of the defect at different initial swimmers' density at $\rho_0=0,1.6\times 10^{-3},2.3\times 10^{-3}, 3.0\times 10^{-3}$, with $v_0 = 0.003$ and $\sigma = 1$. }
(c) The evolution of $D_x$ for different combinations of $v_0$ and $\sigma$ at $\rho_0=3.0\times 10^{-3}$. The black arrows indicate the breakup of the loops. The colored contours represent the superimposed loops at various time snapshots, illustrating their morphodynamics. 
%For the uniform case, particle positions are fixed, with a constant pairwise distance $\sigma = \sqrt[3]{L_x L_y L_z / N_p}$. 
(d) The simulation snapshots showing the midplane colored by the active strength, $\alpha_B(\mathbf{r})$, to indicate the presence of the swimmers. The defect loop is colored according to the twisted angle $\beta$. The black lines represent the director field of the nematic background. The top panel, enclosed in the blue dashed box, corresponds to $v_0 = 0.005$ and $\sigma = 1$ (Supplementary Video 4), while the bottom panel, enclosed in the solid blue box, corresponds to $v_0 = 0.005$ and $\sigma = 3$. The insets show the isosurface of $\alpha_B(\mathbf{r}) = 6 \times 10^{-4}$, highlighting the swimmer cluster.}
    \vspace{\spaceBelowCaption}
    \label{fig:LoopS}
\end{figure*}

\begin{figure*}[htp]
	\centering
	%\vspace{\spaceAboveFigure}
	\includegraphics[width=\linewidth]{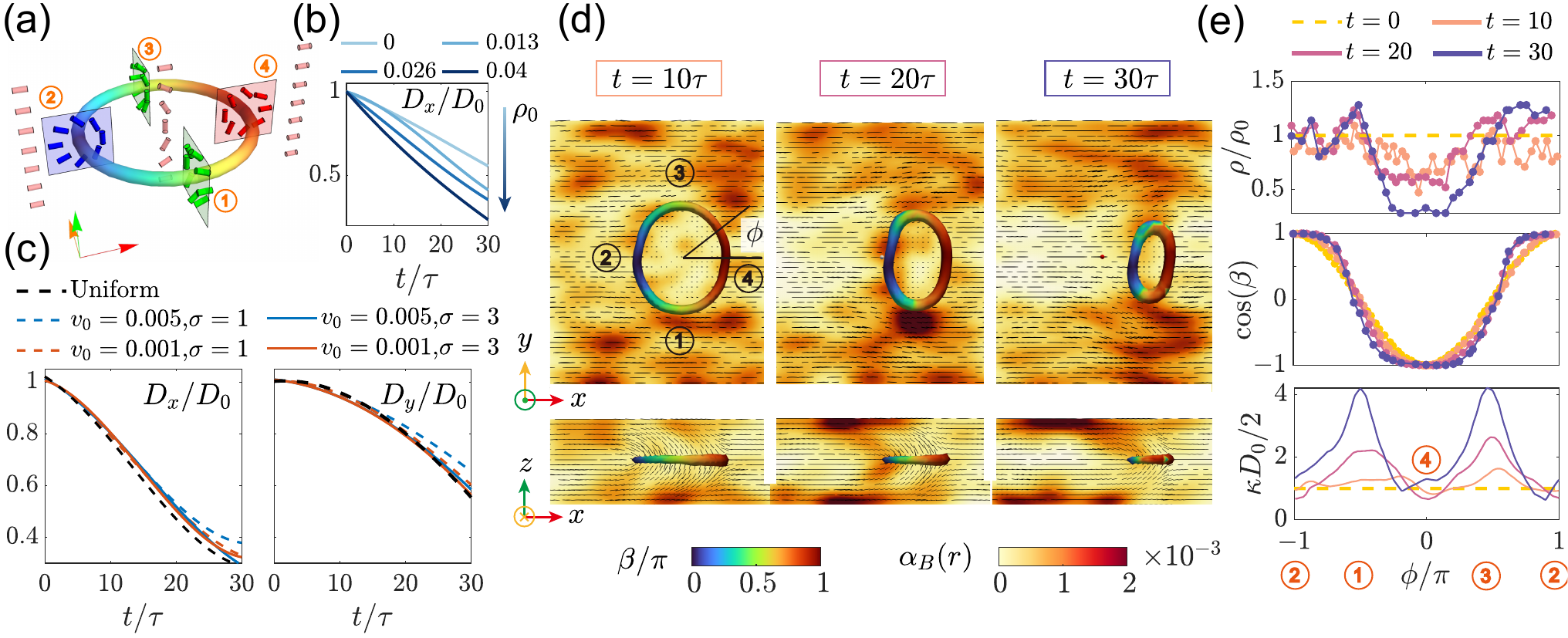}
	\vspace{\spaceBelowFigure}
	\phantomsubfloat{fig:LoopB:a}
    \phantomsubfloat{fig:LoopB:b}
    \phantomsubfloat{fig:LoopB:c}
    \phantomsubfloat{fig:LoopB:d}
    \phantomsubfloat{fig:LoopB:e}
    %\vspace{-2\baselineskip}% Remove extra line inserted by subfloat
    \vspace{-2\baselineskip}
\caption{Dynamics of a wedge-twist loop defect in a planar-anchoring cell \textcolor{black}{with $L_x=45$, $L_y=55$, $L_z=18$, $\zeta_B=0.1$, and $l/\sigma=2$}. (a) Schematic of the wedge-twist loop defect with its $+1/2$ wedge pointing inward. \textcolor{black}{(b) The evolution of $D_x$ at $\rho_0=0,0.013,0.026,0.04$, with $v_0 = 0.005$ and $\sigma = 1$.}
(c) Evolution of the normalized $D_x$ and $D_y$ for different values of $v_0$ and $\sigma$ at $\rho_0=0.026$  . (d) Snapshots at various time intervals to show the evolution of the defect loop and the distribution of the swimmers for $v_0 = 0.005$ and $\sigma = 1$ in (c), colored by $\alpha_B(\mathbf{r})$ in the midplane along the $z$ and $y$ directions. (e) Normalized swimmer density $\rho$, twist angle $\beta$, and loop curvature $\kappa$ plotted along the azimuthal direction of the loop at various time intervals.}
    \vspace{\spaceBelowCaption}
    \label{fig:LoopB}
\end{figure*}

Before we study the interactions between microswimmers and loop defects, we first characterize the structure of the defects, as illustrated in \cref{fig:order_param}. The local director profile of a winding of a defect loop can be written as~\cite{binysh2020}:
\begin{equation}
\mathbf{n}=\cos \frac{1}{2} \phi \mathbf{m}+\sin \frac{1}{2} \phi(\cos \beta \mathbf{m} \times \mathbf{t}+\sin \beta \mathbf{t}),
\end{equation}
where the unit vector $\mathbf{t}$ represents the tangent vector to the defect loop \textcolor{black}{at a specific point. The vector $\pmb\Omega$ defines the axis around which all directors on the cross-sectional plane of the defect loop rotate, and $\beta$ is the angle between $\mathbf{t}$ and $\pmb\Omega$, as shown in \cref{fig:order_param:a,fig:order_param:b}. The vector $\mathbf{m}=\mathbf{t} \times \pmb{\Omega}$ lies in the plane defined by $\mathbf{e}_1$ and $\mathbf{e}_2$ (an orthogonal basis on the plane perpendicular to $\mathbf{t}$). 
By convention, we set $\phi=0$ by $\mathbf{e}_1$. 
The ``phase offset'' angle $\alpha$ is defined as the angle between $\mathbf{m}$ and $\mathbf{e}_1$ ($\mathbf{m}=\cos \alpha \mathbf{e}_1+\sin \alpha \mathbf{e}_2$).} 
A winding with $\beta=\pi/2$ is a pure-twist winding, and those with $\beta=0$ and $\pi$ are called wedges (\cref{fig:order_param:c}).
The offset angle $\alpha$ distinguish the ``twist type'' as ``radial twist'' if $\alpha=0$ or $\pi$ and ``tangential twist'' if $\alpha=\pi/2$ (\cref{fig:order_param:c}).

In the simulation we first study a wedge-twist loop with its winding angle $\beta$ continuously increasing from $0$ to $\pi$ on one half of the loop and decreasing back to $0$ on the other half. The active force will concentrate on and thereby mobilize the $+1/2$ winding (where $\beta=0$) along its orientation (head direction). Therefore, the defect loop may exhibit expansion or shrinkage depending on the orientation of the $+1/2$ wedge.

We first choose the orientation of the $+1/2$ wedge to point outward (i.e., along the $-x$ direction, see \cref{fig:LoopS:a}). This kind of loop defect can be prepared in a nematic cell with homeotropic-anchoring walls, and the nematic director within the plane of the loop is in the $x$-direction (\cref{fig:LoopS:a}). 
\textcolor{black}{We configure the cylindrical region inside the loop to exhibit splay deformation, while the region outside remains uniform. 
We record the lateral size of the defect loop evolution, $D_x$, and the shape evolution of the loop in \cref{fig:LoopS:b,fig:LoopS:c}.
This configuration is inherently unstable. As shown in \cref{fig:LoopS:b}, When no swimmers are present in the system ($\rho_0=0$), the loop rapidly shrinks due to elastic forces. However, as the swimmer population increases, the loop dynamics transition gradually from shrinkage to expansion. At higher swimmer densities, the active flow generated by the microswimmers becomes sufficient to overcome the elastic shrinkage force, stabilizing and growing the defect loop~\cite{kralj2023}. In the subsequent discussions, we focus on a relatively high density system where active flow dominates.
We vary the intrinsic swimming velocity $v_0$ and characteristic size of the swimmers $\sigma$ to explore the effects of the polar nature of the swimmers and their aggregation in 3D LNs.} 
%To highlight the critical role of the polar nature of the swimmers and their aggregation in 3D LNs, we tune two important parameters in the simulation: the intrinsic swimming velocity $v_0$ and the characteristic size of the swimmers $\sigma$ ($l/\sigma=2$). 
\textcolor{black}{The uniform case corresponds to a scenario where the swimmers are evenly distributed throughout the system, maintaining a constant pairwise distance $\sigma = \sqrt[3]{L_x L_y L_z / N_p}$, but their positions remain fixed and cannot evolve. This situation is analogous to active nematic systems described in Refs.~\cite{binysh2020,duclos2020,kralj2023}.} 
The range of swimmer speeds ($0 \leq v_0 \leq 0.005$) is consistent with values observed in biological systems, such as bacterial swimming speeds reported to range from $0-16~\mu$m/s in experiments~\cite{zhou2014,zhou2017,mushenheim2015}. Although the size of individual swimmers is underestimated in our model ($0.2-8~\mu$m from experiments~\cite{zhou2014,zhou2017,mushenheim2015}) due to the point-force assumption—similar to those in Refs.~\cite{genkin2017, genkin2018, turiv2020}—this approach effectively captures the key phenomena identified in the previous section.

%Our simulation resembles the uniform activity system (where the active stress can be expressed as: $\pmb{\Pi}^a=\alpha_\text{uni} \mathbf{Q}$) when the evolution of the positions of the swimmers are restricted by setting $v_0 \to 0$ and removing the background flow coupling in Eqs.~\ref{equ:pos} and \ref{equ:3D_Ori}. The pairwise distance is given by $\sigma = \sqrt[3]{N_x N_y N_z / N_p}$. 

From the colormap of $\alpha_B(\mathbf{r})$ in \cref{fig:LoopS:d}, we observe that the microswimmers concentrate in splay regions, particularly near the $+1/2$ wedge. As a result, the loop is deformed by the active flow, forming a sharp corner pointing toward the $-x$ axis. The increased local curvature of the defect loop further amplifies the splay deformation, thereby attracting more microswimmers. 
In the case of a large swimming velocity and a small swimmer size ($v_0 = 0.005$, $\sigma = 1$), as shown in \cref{fig:LoopS:c} and the upper panel of \cref{fig:LoopS:d}, the particles rapidly aggregate within a small region near the $+1/2$ wedge. The dense clustering of the swimmers generates a larger active flow compared to the uniform activity case, while the depletion of the particles near the $-1/2$ wedge (where $\beta=\pi$) has a minimal effect. Consequently, instead of shrinking as in the uniform activity case, the loop continuously expands. Eventually, the small segment of the loop is dragged so strongly that the loop engenders a child loop moving along the $-x$ with a higher speed, leading to the steady increase of $D_x$ (\cref{fig:LoopS:d}). 
Interestingly, when both the swimming velocity and the repulsive distance are large ($v_0 = 0.005$, $\sigma = 3$), the behavior of the loop changes. In the early stage, the swimmers also rapidly aggregate near the leftmost side of the loop where $\beta\approx0$. However, the larger separations between the swimmers reduce the equivalent activity $\alpha_B$ and weaken the active flow, which cannot compete with the elastic force that tends to shrink the loop. As a result, the swimmer cluster is left ahead of the loop, and the loop annihilates faster than in the uniform activity case.
%Compared to the rounded deformation of the defect loop with an uniform activity, which exhibits a slightly smaller horizontal expansion over the same time interval, the deformation effect is more pronounced around the $+1/2$ defect, coinciding with the region of greater splay. 

A different type of wedge-twist loop with the $+1/2$ wedge pointing inward (along the $+x$ direction, see \cref{fig:LoopB:a}) can be prepared in a nematic cell with planar-anchoring walls. The nematic director within the plane of this type of defect is in the $z$-direction (\cref{fig:LoopB:a}). Therefore, the active force acting on the wedge promotes the shrinking of the defect (\cref{fig:LoopB:b}), which will appear to be elongated in the $y$-direction~\cite{binysh2020}. 
In our simulation, the difference between this type of loop and the above-mentioned loop (\cref{fig:LoopS:a}) is not only the direction of the active force~\cite{binysh2020}, but also the evolution of the swimmer distribution, and the loop structure and its shape (\cref{fig:LoopB:d,fig:LoopB:e} and Supplementary Video 5).
From the snapshots in \cref{fig:LoopB:d} and the azimuthal distribution of the swimmer density $\rho$ (\cref{fig:LoopB:e}), the microswimmers concentrate at the two ends of the loop along the $y$-direction, where the windings are of pure-twist type with $\beta\approx \pi/2$. The shrinking of the loop results in the steady increase of the local curvature of these pure-twist windings (\cref{fig:LoopB:e}). The swimmer density marginally exceeds the initial density $\rho_0$ near the $+1/2$ wedge but continuously decreases around the $-1/2$ wedge.  
When viewed in the $xz$ plane, the swimmers swim toward the two substrates, where splay deformation dominates, and are thereafter trapped, which resembles the trapping of microswimmers in the C-pattern~\cite{turiv2020} (\cref{fig:PolarJet}).  
The shrinking dynamics of the loop does not exhibit significant differences between different sets of $v_0$ and $\sigma$ (\cref{fig:LoopB:c}). However, when the swimming velocity is high while its excluded-volume size is small ($v_0 = 0.005$ and $\sigma = 1$), loop-shrinking dynamics are slightly slowed down, as the swimmers can quickly migrate toward the boundaries and weaken the active flows.

\begin{figure*}[htp]
	\centering
	%\vspace{\spaceAboveFigure}
	\includegraphics[width=\linewidth]{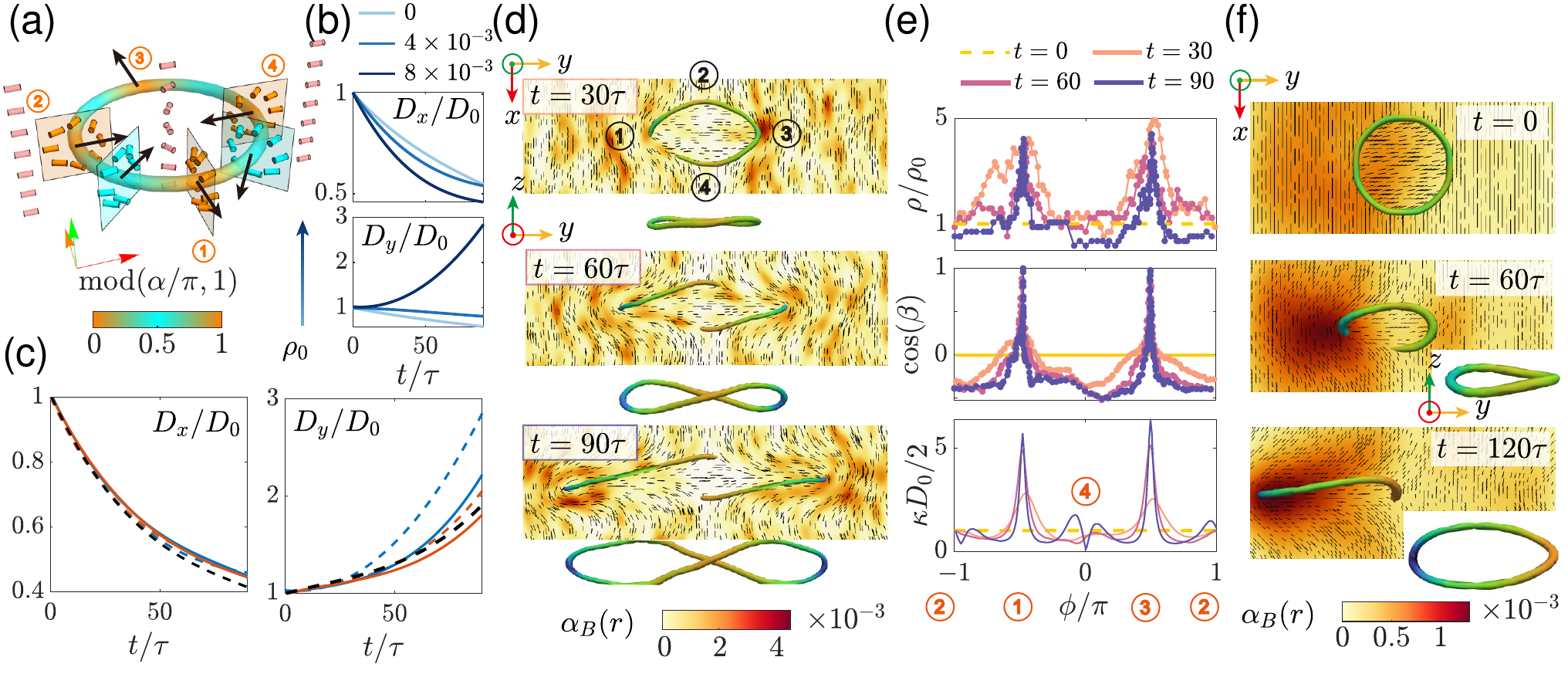}
	\vspace{\spaceBelowFigure}
	\phantomsubfloat{fig:Loopt:a}
    \phantomsubfloat{fig:Loopt:b}
    \phantomsubfloat{fig:Loopt:c}
    \phantomsubfloat{fig:Loopt:d}
    \phantomsubfloat{fig:Loopt:e}
    \phantomsubfloat{fig:Loopt:f}
    %\vspace{-2\baselineskip}% Remove extra line inserted by subfloat
    \vspace{-2\baselineskip}
\caption{Dynamics of a pure-twist loop defect in a planar-anchoring cell \textcolor{black}{with $L_x=80$, $L_y=140$, $L_z=18$, $\zeta_B=0.3$, and $l/\sigma=2$}. (a) Schematic of the pure-twist loop defect. The black arrows show the direction of the active flow if a uniform activity is assumed. \textcolor{black}{(b) The evolution of  $D_x$ and $D_y$ at $\rho_0=0,4.0\times 10^{-3},8.0\times 10^{-3}$, with $v_0 = 0.005$ and $\sigma = 1$.} (c) Evolution of the normalized $D_x$ and $D_y$ for different values of $v_0$ and $\sigma$ when $\rho_0=8.0\times 10^{-3}$. The legends are the same as~\cref{fig:LoopB:c}. Projections of the defect loop in the $xy$ and $yz$ planes, along with the swimmer distribution in the midplane, for $v_0 = 0.005$ and $\sigma = 1$ in (c). (e) A pure-twist loop surrounded by the same director configuration as in (f) but with an initially nonuniform distribution of the swimmers with total population $N_p=600$ ($v_0 = 0.002$ and $\sigma = 3$). The left side of the simulation box ($y < N_y/2$) contains triple as many microswimmers as the right side ($y \geq N_y/2$). The dynamics of the loop are demonstrated through changes in its position and structure, as characterized by the angle $\beta$.}
    \vspace{\spaceBelowCaption}
    \label{fig:Loopt}
\end{figure*}
%the shrinking is slowed down when $v_0 = 0.005$ and $\sigma = 1$, as bacteria swim outward along the $\pm y$ and $\pm z$ directions and rapidly adhere to the boundaries. This reduces the total number of the swimmers around the loop, thereby slightly affecting the shrinking dynamics.
%We plot the normalized bacterial density $\rho$, the twist angle $\beta$ and the loop curvature $\kappa$ along the azimuthal direction of the loop at different time in \cref{fig:LoopB:d}. 

We further study pure-twist loops in our simulation. For a pure-twist loop, $\beta=\pi/2$ everywhere and $\alpha$ continuously winds $4\pi$ along the loop (\cref{fig:Loopt:a}). 
For a uniform activity, the loop expands at position 1 and 3, where the windings are of radial twist type with the active flow pointing outward, and shrinks at position 2 and 4, where the windings are of radial twist type with active flow pointing inward. The out-of-plane flow at other windings of the loop causes the loop to buckle at its four diagonal directions~\cite{binysh2020}. 
\textcolor{black}{In \cref{fig:Loopt:b}, an increase in $\rho_0$ results in a transition from shrinkage to expansion of the loop in the $y$-direction. The faster expansion in the $y$-direction leads to a corresponding faster shrinkage in the $x$-direction.} From \cref{fig:Loopt:c,fig:Loopt:d,fig:Loopt:e} and Supplementary Video 6, the swimmers gradually concentrate at position 1 and 3 where $\beta$ angle decreases over time, and their curvatures increase as the loop becomes stretched along the $y$-direction. Loop elongation $D_y$ exhibits the fastest growth rate for high swimming velocity and small swimmer size ($v_0 = 0.005$ and $\sigma = 1$), whereas its shrinkage along the $x$-direction is less sensitive to these parameters.
Motivated by the fact that swimmer density variation around the defect loop can impact its dynamics, we create a nonuniform initial distribution of the swimmers in the simulation box, with the $+y$ half of the box populated with $ 3N_p/4$ swimmers and the $-y$ half of the box with $N_p/4$ swimmers (\cref{fig:Loopt:f}). 
In the high-density region, the bacteria migrate to position 1 over time, and the loop undergoes a torsional buckling, causing the immersed portion of the loop to be drawn along the $-y$-direction and twisted significantly. Meanwhile, swimmers in the $+y$ half of the box initially aggregate around position 3 at $t = 60\tau$. However, this small cluster of swimmers are unable to stretch the loop along the $+y$-direction and are eventually dispersed ($t = 120\tau$ in \cref{fig:Loopt:f}). 
As a result, the defect loop undergoes a net displacement in the $-y$ direction. The unbalanced torsional twist in the $+y$ and $-y$ halves of the loop causes it to eventually tumble into the $yz$ plane, transforming the loop from an initially pure-twist configuration into a wedge-twist configuration. This new configuration exhibits a similar nematic structure to the wedge-twist loop in the homeotropic cell (\cref{fig:Loopt:f}, \cref{fig:LoopS:a}).

%significant torsional twist in the $-y$ portion is followed by elastic relaxation in the $+y$ half, ultimately causing the loop to unbuckle and settle into the $yz$ plane. 
%Simultaneously, a transformation occurs from a pure-twist type to a wedge-twist type, with the rotation vector $\pmb{\Omega}$ consistently pointing along the positive $z$ axis before and after the transformation. At $t = 120\tau$, the director field resembles that shown in \cref{fig:LoopS:a}, with a rotation of the coordinate axes: $x \to y$, $z \to -x$, and $y \to z$. 
Our results reveal that not only do the local profiles of the defect loops influence their dynamics in an LN, but also the distribution of the active units plays a crucial role. Their interplay can give rise to rich dynamics and lead to different destinies of the defects. 

%The three types of defect loops described above are confined in a box with periodic boundary conditions, where bacteria penetrate the bounding surfaces at homeotropically anchored boundaries and align horizontally at planar anchored boundaries. This setup provides fundamental insights into how bacteria aggregate and deform a defect loop in the bulk. Specifically, regions with higher bacterial density consistently coincide with areas of high curvature and $+1/2$ wedge-type defects, while bacteria are depleted around segments with lower curvature and $-1/2$ wedge-type defects. This correlation holds true regardless of the type of loop or the bounding conditions. 
%Next, we consider a more realistic scenario where the bounding surfaces in the $z$ direction are non-penetrable, and that the bacteria can reverse their swimming directions. 

\subsection{Density Pumping in a Hybrid Nematic Cell}
In the above discussions, the swimmers can re-enter the simulation box via periodic boundary conditions. This simplification has helped us to focus on the dynamics of the defects immersed in an environment with a constant overall microswimmer density. We now focus on a more realistic situation that the microswimmers are allowed to aggregate on boundary walls. It is observed that bacteria can switch their swimming direction by $180^\degree$ in an anisotropic environment, with their flagella switching sides by backtracking or forming two opposing tails to perform a ``tug-of-oars'' mechanism~\cite{patteson2015, mushenheim2015, zhou2017, goral2022, prabhune2024}. This behavior arises due to the suppression of the run-and-tumble motion typically observed in isotropic solutions~\cite{berg1972, taute2015, Figueroa-Morales2020}. 
For an LC environment, the reversal time $t_\textnormal{rev}$ varies depending on the type and size of the bacterium, the concentration of the LC solution, and the anchoring type and strength of the boundary surfaces. 
For instance, $t_\textnormal{rev} \sim 30$ $-$ $\SI{300}{\second}$ for \textit{B. subtilis} in lyotropic disodium cromoglycate (DSCG) over a wide range of concentrations (\SI{12.5}{wt.\percent}–\SI{18.5}{wt.\percent})~\cite{genkin2017,prabhune2024}; and $t_\textnormal{rev} \sim$ \SI{1e4}{\second} for \textit{B. subtilis} in DSCG solutions with patterned anchoring on the top and bottom surfaces~\cite{zhou2017}; 
$t_\textnormal{rev} \sim $ \SI{7}{\second} for \textit{P. mirabilis} in the bulk of \SI{15}{wt.\percent} DSCG solution with homeotropic or hybrid surface anchoring~\cite{mushenheim2015}; Moreover, \textit{B. subtilis} can swim horizontally near the homeotropically anchored surfaces and switch their swimming direction either by backtracking or through an horizontal-vertical-horizontal mechanism with $t_\textnormal{rev} \sim$ 2 $-$ \SI{12}{\second} in a nematic DSCG solution~\cite{zhou2017}. 
\begin{figure}[htp]
	\centering
	%\vspace{\spaceAboveFigure}
	\includegraphics[width=\linewidth]{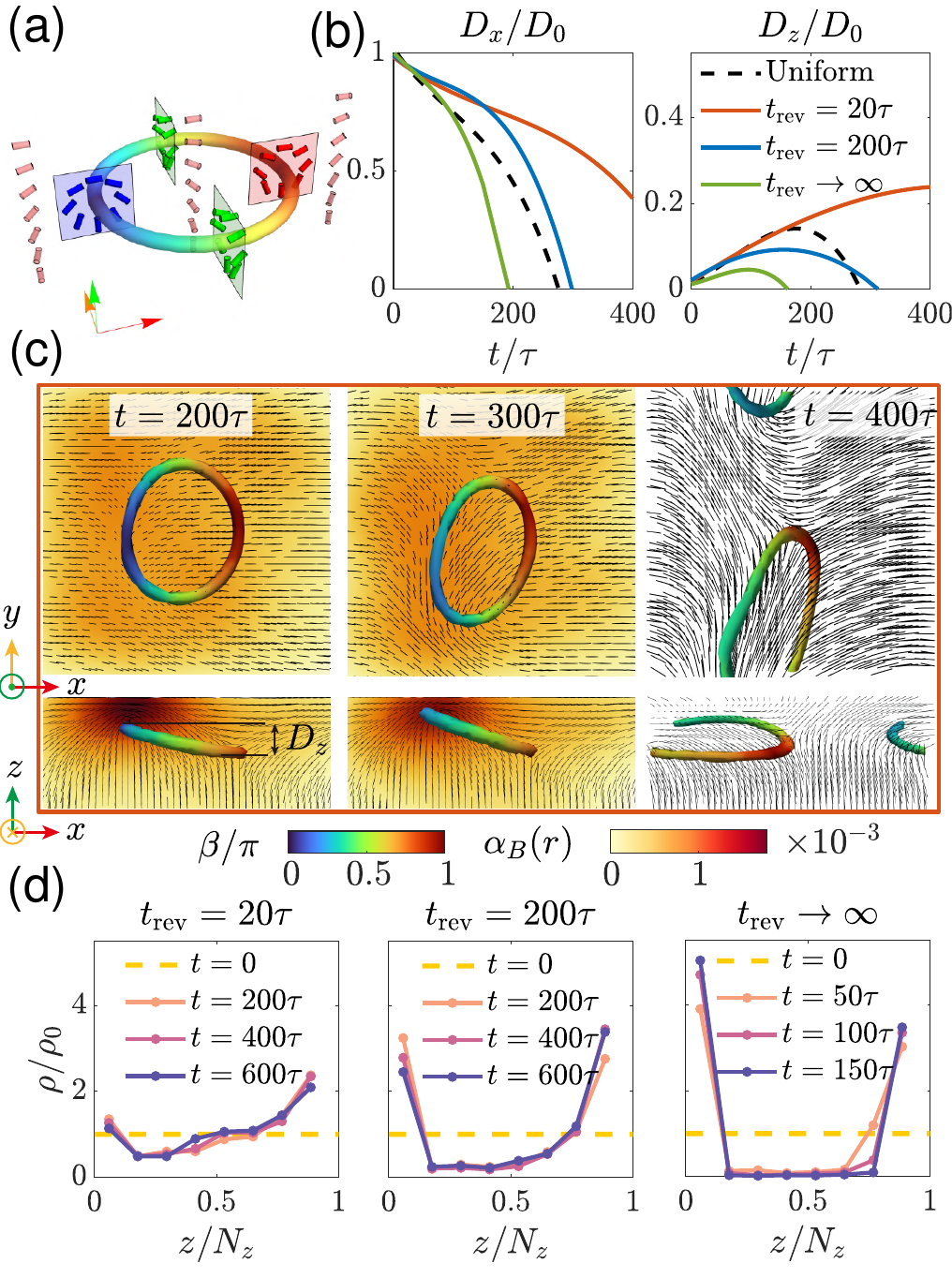}
	\vspace{\spaceBelowFigure}
	\phantomsubfloat{fig:LoopH:a}
    \phantomsubfloat{fig:LoopH:b}
    \phantomsubfloat{fig:LoopH:c}
    \phantomsubfloat{fig:LoopH:d}
    %\vspace{-2\baselineskip}% Remove extra line inserted by subfloat
    \vspace{-1.5\baselineskip}
\caption{Interactions of microswimmers with a wedge-twist loop in a hybrid cell with non-penetrable bounding surfaces with \textcolor{black}{$L_x=45$, $L_y=45$, $L_z=18$, $\zeta_B=0.2$,} $v_0=0.003$, $\sigma=3$, and $l/\sigma=2$.  (a) The schematic of a wedge-twist loop defect in a hybrid cell with its $+1/2$ wedge pointing upward. (b) The evolution of the lateral size, $D_x$ and $D_z$, is analyzed for a range of $t_\mathrm{rev}$ spanning from $0$ (uniform) to $\infty$. (c) Snapshots at various time intervals to show the evolution of the loop defect and the distribution of the swimmers for $t_\mathrm{rev}=20\tau$. (d) The density distribution along the $z$ direction is shown for $t_\mathrm{rev} = 20\tau$, $t_\mathrm{rev} = 200\tau$, and $t_\mathrm{rev} \to \infty$. For the uniform case, note that $\rho / \rho_0$ remains consistently equal to 1.}
    \vspace{\spaceBelowCaption}
    \label{fig:LoopH}
\end{figure}

In our simulation, the swimming direction switches as $\mathbf{p}_i \to -\mathbf{p}_i$ at $t = t_{n,i}$, where the time interval between two consecutive switches, $\Delta T_{n,i} \equiv t_{n,i} - t_{n-1,i}$, is given by $\Delta T_{n,i} = -{\ln(U)}/{m}$. Here, $U$ is a random number in the range $[0,1)$, and the switching of direction is modeled as independent events occurring with an average rate $m = {1}/{t_\mathrm{rev}}$.  When $t_\mathrm{rev} \to 0$, the system reduces to a uniform active nematic~\cite{genkin2017}. Conversely, for $t_\mathrm{rev} \to \infty$, swimmers gradually adhere to the bounding surfaces after starting from a uniformly distributed initial state. 

Here we consider a hybrid wall with the top and bottom surface imposing planar and homeotropic anchoring condition, respectively.
The $+1/2$ wedge on a wedge-twist loop in this type of cell points towards the $+z$ direction (\cref{fig:LoopH:a}).
%By applying hybrid anchoring on the top and bottom surfaces and adjusting the tilting direction along the $z$ axis to be opposite within and outside the cylindrical boundary, we modify the orientation of the $\pm 1/2$ defects, as illustrated in \cref{fig:LoopH:a}. 
The loop size shows non-monotonic behavior as $t_\mathrm{rev}$ changes. 
If $t_\mathrm{rev} \to 0$ (uniform activity), the vertical active flow around the $+1/2$ wedge pins the loop segment and lifts this piece along the $+z$ direction, causing the loop to title out of the $xy$ plane. This leads to a decrease in $D_x$ and an initial increase in $D_z$. However, as the lifting of the $+1/2$ wedge is stopped by the top confinement surface, elastic relaxation begins to dominate, resulting in a decrease in $D_z$ (\cref{fig:LoopH:b}). 
As $t_\mathrm{rev}$ increases to $20 \tau$ ($\approx \SI{2.5}{\second}$), the shrinking process slows down. During this phase, many bacteria are pumped into the upper region of the cell (\cref{fig:LoopH:c,fig:LoopH:d}, see Supplementary Video 7), congregating around the $+1/2$ wedge. This triggers a bend instability, which prevents the annihilation of the defect loop. Consequently, $D_x$ does not approach zero, and $D_z$ continues to increase over time. 
At even higher $t_\mathrm{rev}$ values (e.g., $t_\mathrm{rev} = 200 \tau \approx \SI{25}{\second}$), more swimmers accumulate near the bounding surfaces ($z/N_z = 0, 1$ in \cref{fig:LoopH:d}), waiting for their next reversal to re-enter the bulk. This reduces the population density of the \textcolor{black}{swimmers} in the bulk, and the shrinking of the loop is therefore the fastest. Finally, if $t_\mathrm{rev} \to \infty$, the \textcolor{black}{swimmers} initially accumulate in the upper region of the cell around the $+1/2$ wedge. However, most of them eventually make a U-turn at the head of the $+1/2$ defect and migrate toward the bottom surface, leading to their accumulation at the bottom surface, which exhibits a higher concentration than that near the top surface. This implies that microswimmers could accumulate in homeotropic-anchoring walls where bend deformation dominates. As a consequence of the depletion of bacteria in the bulk at the early stage ($t < 150 \tau$ in \cref{fig:LoopH:d}), the loop shrinks the fastest and is lifted by the shortest distance. 
%While the proportion and likelihood of the swimmer switching their swimming direction may be overestimated in our analysis, we highlight an important factor: the reversal time, which plays a crucial role in understanding the 3D navigation of bacteria.
Further experimental validation is needed to investigate bacterial reversal near different types of anchoring surfaces and to test how this influences disclination lines or loops in the bulk.

\subsection{Discussion}
We have developed a hybrid method to combine a particle-based approach for the microswimmers and a well-validated hydrodynamic model for the continuous LC phase. Our method can work with both 2D and 3D systems.
A similar hybrid approach has been shown successful in simulating the dynamics of active chemotactic droplets dispersed in a chemical field~\cite{chan2024interplay}.
%Our simulation results demonstrate the effectiveness of the hybrid method, combining an agent-based approach for microswimmers with a well-validated hydrodynamic model governing the director field and flow field in both 2D and 3D systems. 
%The motion of the microswimmers is coupled with the director and flow fields; conversely, the superposition of active flows from individual microswimmers modifies the flow field, subsequently distorting the director field. 
This method conveniently models uniaxial swimmers (which is characterized by the shape parameter $e$) as force dipoles. More sophisticated models, such as spherical squirmers, have been used in literature~\cite{lintuvuori2017,gautam2024}. Our method will become more advantageous when the microswimmers are small and their anchoring is not essential or when the population of the microswimmers is enormous. Compared to other field models that treat the bacteria as either another continuous phase~\cite{turiv2020} or a concentration field~\cite{genkin2017, genkin2018, turiv2020}, our method treats the microswimmers as discrete particles without doing mean-field averages, therefore retaining the particle-level resolution and exhibiting robustness and numerical stability over a wide range of microswimmer populations. 
This allows us to examine how micro\textcolor{black}{swimmers} transition from individual dynamics to collective dynamics as their population density increases.
Because our method accurately accounts for the hydrodynamics of the background LC, we are able to reproduce the stabilized undulations of jets and swirls reported in the experiments.
%Additionally, it offers a clear physical interpretation with fewer controlling parameters, thereby reducing the complexity associated with the interface introduction in the two-phase field model~\cite{turiv2020}.
%Notably, this method demonstrates robust performance across diverse microswimmers population variations, capturing the transition from individual to collective behavior and even stabilizing undulation patterns where the hydrodynamic field plays a crucial role — a feature absent in the previous hybrid model~\cite{koizumi2020}. 
%Unlike the previous agent-based model~\cite{koizumi2020}, in our simulation, bacterial locomotion arises from both self-propulsion and advection by the background flow. The influence of bacteria on the nematic field is not mediated by the short-range force localized around the bacteria. Instead, it operates through a long-range velocity field, allowing for the observation of bend instability and nematic field distortions even in regions devoid of bacteria.
%Moreover, the model excels in detailing microswimmers behaviors, yielding results that are more realistic compared to those produced by the advection-diffusion model~\cite{genkin2017, genkin2018, turiv2020}.
\textcolor{black}{Although the LBM method we used to simulate nematic flow does not incorporate thermal fluctuations present in particle-based approaches~\cite{kozhukhov2022,macias2023,mandal2021,kozhukhov2024,mandal2025}, the system still exhibits fluctuations arising from the noise intrinsic to the particle dynamics. 
These fluctuations %, primarily generated through active stresses, 
can induce perturbations in both the nematic flow and the defect structure.} 

Furthermore, the extension to 3D systems enables comparisons to more realistic experimental studies, including the effects of surface anchoring strength, hydrodynamic boundary condition, and 3D nematic field and confinement. 
\textcolor{black}{Although the loop dynamics differ only quantitatively between a uniform active nematic system and the living nematic system with polar swimmers, we find that the active stress induced by the swimmers can be spatially modulated by the nematic structure, as the swimmers prefer to migrate into splay regions, such as wall regions with splay deformations and certain regions around a loop defect; for swimmers with long swimming-direction reversal time, however, they may also tend to accumulate in boundaries with bend deformations. There is a positive feedback effect between the swimmer density and the nematic structure, because active stress can also further deform the nematic director field. Our findings highlight the potential of living nematics to control the 3D structure of a nematic system through nonuniformly distributed microswimmers, as well as to manipulate the 3D spatial distribution of microswimmers using a nematic environment.} 
There are many potential applications of our work in extended areas, including the study of how anisotropic elastic constants affect the dynamics of the microswimmers, microswimmers interacting with loop defects with nonzero topological charge~\cite{copar2019,carenza2019} or solitonic structures in cholesteric LCs~\cite{whitfield2017,carenza2019}, interactions between microswimmers in active turbulent state~\cite{duclos2020,kralj2023}, impact of external fields~\cite{boule2020}.

\begin{acknowledgments}
R.Z. acknowledges support from Hong Kong Research Grants Council via grant no. 16306924. Z.Y. acknowledges support from National Natural Science Foundation of China No. 12374219.
\end{acknowledgments}

\begin{flushleft}
\subsection{Supplementary Information}
\textbf{Supplementary Video 1}: Steady circulation and subsequent undulation of bacteria in a spiral-patterned thin film, as shown in \cref{fig:spiral}.

\textbf{Supplementary Videos 2 \& 3}: Stabilized undulation observed in the splay region for \(N_p = 220\) and \(N_p = 280\), respectively, as illustrated in \cref{fig:PolarJet}. 
\textcolor{black}{The bottom video illustrates the measurement process. Swimmers near the centerline are represented as blue dots, while the thick red centerline depicts the shape of the jet, with its length denoted as $C_l$. Green dashed lines, drawn perpendicular to the centerline, indicate the distances of the swimmers from the centerline, which are used to calculate the jet's width $W_d$. Pink triangles mark the peak positions of the wave.}

\textbf{Supplementary Video 4}: Dynamics of bacteria and loops within a cylindrical splay cell, corresponding to \cref{fig:LoopS}.

\textbf{Supplementary Videos 5 \& 6}: Dynamics of bacteria and loops within cylindrical bend and cylindrical twist cells, respectively, as depicted in \cref{fig:LoopB}.

\textbf{Supplementary Video 7}: Dynamics of bacteria and loops in a cylindrical reversed hybrid cell as shown in \cref{fig:LoopH}.
\end{flushleft}

\clearpage
\bibliographystyle{jabbrv_apsrev4-2}
\bibliography{Bacteria}

\begin{thebibliography}{77}%
\makeatletter
\providecommand \@ifxundefined [1]{%
 \@ifx{#1\undefined}
}%
\providecommand \@ifnum [1]{%
 \ifnum #1\expandafter \@firstoftwo
 \else \expandafter \@secondoftwo
 \fi
}%
\providecommand \@ifx [1]{%
 \ifx #1\expandafter \@firstoftwo
 \else \expandafter \@secondoftwo
 \fi
}%
\providecommand \natexlab [1]{#1}%
\providecommand \enquote  [1]{``#1''}%
\providecommand \bibnamefont  [1]{#1}%
\providecommand \bibfnamefont [1]{#1}%
\providecommand \citenamefont [1]{#1}%
\providecommand \href@noop [0]{\@secondoftwo}%
\providecommand \href [0]{\begingroup \@sanitize@url \@href}%
\providecommand \@href[1]{\@@startlink{#1}\@@href}%
\providecommand \@@href[1]{\endgroup#1\@@endlink}%
\providecommand \@sanitize@url [0]{\catcode `\\12\catcode `\$12\catcode
  `\&12\catcode `\#12\catcode `\^12\catcode `\_12\catcode `\%12\relax}%
\providecommand \@@startlink[1]{}%
\providecommand \@@endlink[0]{}%
\providecommand \url  [0]{\begingroup\@sanitize@url \@url }%
\providecommand \@url [1]{\endgroup\@href {#1}{\urlprefix }}%
\providecommand \urlprefix  [0]{URL }%
\providecommand \Eprint [0]{\href }%
\providecommand \doibase [0]{https://doi.org/}%
\providecommand \selectlanguage [0]{\@gobble}%
\providecommand \bibinfo  [0]{\@secondoftwo}%
\providecommand \bibfield  [0]{\@secondoftwo}%
\providecommand \translation [1]{[#1]}%
\providecommand \BibitemOpen [0]{}%
\providecommand \bibitemStop [0]{}%
\providecommand \bibitemNoStop [0]{.\EOS\space}%
\providecommand \EOS [0]{\spacefactor3000\relax}%
\providecommand \BibitemShut  [1]{\csname bibitem#1\endcsname}%
\let\auto@bib@innerbib\@empty
%</preamble>
\bibitem [{\citenamefont {Marchetti}\ \emph {et~al.}(2013)\citenamefont
  {Marchetti}, \citenamefont {Joanny}, \citenamefont {Ramaswamy}, \citenamefont
  {Liverpool}, \citenamefont {Prost}, \citenamefont {Rao},\ and\ \citenamefont
  {Simha}}]{marchetti2013}%
  \BibitemOpen
  \bibfield  {author} {\bibinfo {author} {\bibfnamefont {M.~C.}\ \bibnamefont
  {Marchetti}}, \bibinfo {author} {\bibfnamefont {J.-F.}\ \bibnamefont
  {Joanny}}, \bibinfo {author} {\bibfnamefont {S.}~\bibnamefont {Ramaswamy}},
  \bibinfo {author} {\bibfnamefont {T.~B.}\ \bibnamefont {Liverpool}}, \bibinfo
  {author} {\bibfnamefont {J.}~\bibnamefont {Prost}}, \bibinfo {author}
  {\bibfnamefont {M.}~\bibnamefont {Rao}},\ and\ \bibinfo {author}
  {\bibfnamefont {R.~A.}\ \bibnamefont {Simha}},\ }\bibfield  {title} {\bibinfo
  {title} {\emph {Hydrodynamics of soft active matter}},\ }\href
  {https://doi.org/https://doi.org/10.1103/RevModPhys.85.1143} {\bibfield
  {journal} {\bibinfo  {journal} {\protect\JournalTitle{Reviews of modern
  physics}}\ }\textbf {\bibinfo {volume} {85}},\ \bibinfo {pages} {1143}
  (\bibinfo {year} {2013})}\BibitemShut {NoStop}%
\bibitem [{\citenamefont {Saintillan}(2018)}]{saintillan2018}%
  \BibitemOpen
  \bibfield  {author} {\bibinfo {author} {\bibfnamefont {D.}~\bibnamefont
  {Saintillan}},\ }\bibfield  {title} {\bibinfo {title} {\emph {Rheology of
  active fluids}},\ }\href
  {https://doi.org/https://doi.org/10.1146/annurev-fluid-010816-060049}
  {\bibfield  {journal} {\bibinfo  {journal} {\protect\JournalTitle{Annual
  review of fluid mechanics}}\ }\textbf {\bibinfo {volume} {50}},\ \bibinfo
  {pages} {563} (\bibinfo {year} {2018})}\BibitemShut {NoStop}%
\bibitem [{\citenamefont {Gompper}\ \emph {et~al.}(2020)\citenamefont
  {Gompper}, \citenamefont {Winkler}, \citenamefont {Speck}, \citenamefont
  {Solon}, \citenamefont {Nardini}, \citenamefont {Peruani}, \citenamefont
  {L{\"o}wen}, \citenamefont {Golestanian}, \citenamefont {Kaupp},
  \citenamefont {Alvarez} \emph {et~al.}}]{gompper2020}%
  \BibitemOpen
  \bibfield  {author} {\bibinfo {author} {\bibfnamefont {G.}~\bibnamefont
  {Gompper}}, \bibinfo {author} {\bibfnamefont {R.~G.}\ \bibnamefont
  {Winkler}}, \bibinfo {author} {\bibfnamefont {T.}~\bibnamefont {Speck}},
  \bibinfo {author} {\bibfnamefont {A.}~\bibnamefont {Solon}}, \bibinfo
  {author} {\bibfnamefont {C.}~\bibnamefont {Nardini}}, \bibinfo {author}
  {\bibfnamefont {F.}~\bibnamefont {Peruani}}, \bibinfo {author} {\bibfnamefont
  {H.}~\bibnamefont {L{\"o}wen}}, \bibinfo {author} {\bibfnamefont
  {R.}~\bibnamefont {Golestanian}}, \bibinfo {author} {\bibfnamefont {U.~B.}\
  \bibnamefont {Kaupp}}, \bibinfo {author} {\bibfnamefont {L.}~\bibnamefont
  {Alvarez}}, \emph {et~al.},\ }\bibfield  {title} {\bibinfo {title} {\emph
  {The 2020 motile active matter roadmap}},\ }\href
  {https://doi.org/https://iopscience.iop.org/article/10.1088/1361-648X/ab6348/meta}
  {\bibfield  {journal} {\bibinfo  {journal} {\protect\JournalTitle{Journal of
  Physics: Condensed Matter}}\ }\textbf {\bibinfo {volume} {32}},\ \bibinfo
  {pages} {193001} (\bibinfo {year} {2020})}\BibitemShut {NoStop}%
\bibitem [{\citenamefont {Bechinger}\ \emph {et~al.}(2016)\citenamefont
  {Bechinger}, \citenamefont {Di~Leonardo}, \citenamefont {L{\"o}wen},
  \citenamefont {Reichhardt}, \citenamefont {Volpe},\ and\ \citenamefont
  {Volpe}}]{bechinger2016}%
  \BibitemOpen
  \bibfield  {author} {\bibinfo {author} {\bibfnamefont {C.}~\bibnamefont
  {Bechinger}}, \bibinfo {author} {\bibfnamefont {R.}~\bibnamefont
  {Di~Leonardo}}, \bibinfo {author} {\bibfnamefont {H.}~\bibnamefont
  {L{\"o}wen}}, \bibinfo {author} {\bibfnamefont {C.}~\bibnamefont
  {Reichhardt}}, \bibinfo {author} {\bibfnamefont {G.}~\bibnamefont {Volpe}},\
  and\ \bibinfo {author} {\bibfnamefont {G.}~\bibnamefont {Volpe}},\ }\bibfield
   {title} {\bibinfo {title} {\emph {Active particles in complex and crowded
  environments}},\ }\href
  {https://doi.org/https://doi.org/10.1103/RevModPhys.88.045006} {\bibfield
  {journal} {\bibinfo  {journal} {\protect\JournalTitle{Reviews of modern
  physics}}\ }\textbf {\bibinfo {volume} {88}},\ \bibinfo {pages} {045006}
  (\bibinfo {year} {2016})}\BibitemShut {NoStop}%
\bibitem [{\citenamefont {Chan}\ \emph
  {et~al.}(2024{\natexlab{a}})\citenamefont {Chan}, \citenamefont {Wu},
  \citenamefont {Qiao}, \citenamefont {Fong}, \citenamefont {Yang},
  \citenamefont {Han},\ and\ \citenamefont {Zhang}}]{chan2024}%
  \BibitemOpen
  \bibfield  {author} {\bibinfo {author} {\bibfnamefont {C.~W.}\ \bibnamefont
  {Chan}}, \bibinfo {author} {\bibfnamefont {D.}~\bibnamefont {Wu}}, \bibinfo
  {author} {\bibfnamefont {K.}~\bibnamefont {Qiao}}, \bibinfo {author}
  {\bibfnamefont {K.~L.}\ \bibnamefont {Fong}}, \bibinfo {author}
  {\bibfnamefont {Z.}~\bibnamefont {Yang}}, \bibinfo {author} {\bibfnamefont
  {Y.}~\bibnamefont {Han}},\ and\ \bibinfo {author} {\bibfnamefont
  {R.}~\bibnamefont {Zhang}},\ }\bibfield  {title} {\bibinfo {title} {\emph
  {Chiral active particles are sensitive reporters to environmental
  geometry}},\ }\href
  {https://doi.org/https://doi.org/10.1038/s41467-024-45531-5} {\bibfield
  {journal} {\bibinfo  {journal} {\protect\JournalTitle{Nature
  Communications}}\ }\textbf {\bibinfo {volume} {15}},\ \bibinfo {pages} {1406}
  (\bibinfo {year} {2024}{\natexlab{a}})}\BibitemShut {NoStop}%
\bibitem [{\citenamefont {Suarez}\ and\ \citenamefont
  {Pacey}(2006)}]{suarez2006}%
  \BibitemOpen
  \bibfield  {author} {\bibinfo {author} {\bibfnamefont {S.~S.}\ \bibnamefont
  {Suarez}}\ and\ \bibinfo {author} {\bibfnamefont {A.}~\bibnamefont {Pacey}},\
  }\bibfield  {title} {\bibinfo {title} {\emph {Sperm transport in the female
  reproductive tract}},\ }\href
  {https://doi.org/https://doi.org/10.1093/humupd/dmi047} {\bibfield  {journal}
  {\bibinfo  {journal} {\protect\JournalTitle{Human reproduction update}}\
  }\textbf {\bibinfo {volume} {12}},\ \bibinfo {pages} {23} (\bibinfo {year}
  {2006})}\BibitemShut {NoStop}%
\bibitem [{\citenamefont {Kantsler}\ \emph {et~al.}(2014)\citenamefont
  {Kantsler}, \citenamefont {Dunkel}, \citenamefont {Blayney},\ and\
  \citenamefont {Goldstein}}]{kantsler2014}%
  \BibitemOpen
  \bibfield  {author} {\bibinfo {author} {\bibfnamefont {V.}~\bibnamefont
  {Kantsler}}, \bibinfo {author} {\bibfnamefont {J.}~\bibnamefont {Dunkel}},
  \bibinfo {author} {\bibfnamefont {M.}~\bibnamefont {Blayney}},\ and\ \bibinfo
  {author} {\bibfnamefont {R.~E.}\ \bibnamefont {Goldstein}},\ }\bibfield
  {title} {\bibinfo {title} {\emph {Correction: Rheotaxis facilitates upstream
  navigation of mammalian sperm cells}},\ }\href
  {https://doi.org/https://doi.org/10.7554/eLife.03521} {\bibfield  {journal}
  {\bibinfo  {journal} {\protect\JournalTitle{Elife}}\ }\textbf {\bibinfo
  {volume} {3}},\ \bibinfo {pages} {e03521} (\bibinfo {year}
  {2014})}\BibitemShut {NoStop}%
\bibitem [{\citenamefont {Smalyukh}\ \emph {et~al.}(2008)\citenamefont
  {Smalyukh}, \citenamefont {Butler}, \citenamefont {Shrout}, \citenamefont
  {Parsek},\ and\ \citenamefont {Wong}}]{smalyukh2008}%
  \BibitemOpen
  \bibfield  {author} {\bibinfo {author} {\bibfnamefont {I.~I.}\ \bibnamefont
  {Smalyukh}}, \bibinfo {author} {\bibfnamefont {J.}~\bibnamefont {Butler}},
  \bibinfo {author} {\bibfnamefont {J.~D.}\ \bibnamefont {Shrout}}, \bibinfo
  {author} {\bibfnamefont {M.~R.}\ \bibnamefont {Parsek}},\ and\ \bibinfo
  {author} {\bibfnamefont {G.~C.~L.}\ \bibnamefont {Wong}},\ }\bibfield
  {title} {\bibinfo {title} {\emph {Elasticity-mediated nematiclike bacterial
  organization in model extracellular DNA matrix}},\ }\href
  {https://doi.org/10.1103/PhysRevE.78.030701} {\bibfield  {journal} {\bibinfo
  {journal} {\protect\JournalTitle{Phys. Rev. E}}\ }\textbf {\bibinfo {volume}
  {78}},\ \bibinfo {pages} {030701} (\bibinfo {year} {2008})}\BibitemShut
  {NoStop}%
\bibitem [{\citenamefont {Flemming}\ and\ \citenamefont
  {Wingender}(2010)}]{flemming2010}%
  \BibitemOpen
  \bibfield  {author} {\bibinfo {author} {\bibfnamefont {H.-C.}\ \bibnamefont
  {Flemming}}\ and\ \bibinfo {author} {\bibfnamefont {J.}~\bibnamefont
  {Wingender}},\ }\bibfield  {title} {\bibinfo {title} {\emph {The biofilm
  matrix}},\ }\href {https://doi.org/https://doi.org/10.1038/nrmicro2415}
  {\bibfield  {journal} {\bibinfo  {journal} {\protect\JournalTitle{Nature
  reviews microbiology}}\ }\textbf {\bibinfo {volume} {8}},\ \bibinfo {pages}
  {623} (\bibinfo {year} {2010})}\BibitemShut {NoStop}%
\bibitem [{\citenamefont {Lemon}\ \emph {et~al.}(2017)\citenamefont {Lemon},
  \citenamefont {Yang}, \citenamefont {Srivastava}, \citenamefont {Luk},\ and\
  \citenamefont {Garza}}]{lemon2017}%
  \BibitemOpen
  \bibfield  {author} {\bibinfo {author} {\bibfnamefont {D.~J.}\ \bibnamefont
  {Lemon}}, \bibinfo {author} {\bibfnamefont {X.}~\bibnamefont {Yang}},
  \bibinfo {author} {\bibfnamefont {P.}~\bibnamefont {Srivastava}}, \bibinfo
  {author} {\bibfnamefont {Y.-Y.}\ \bibnamefont {Luk}},\ and\ \bibinfo {author}
  {\bibfnamefont {A.~G.}\ \bibnamefont {Garza}},\ }\bibfield  {title} {\bibinfo
  {title} {\emph {Polymertropism of rod-shaped bacteria: movement along aligned
  polysaccharide fibers}},\ }\href {https://doi.org/10.1038/s41598-017-07486-0}
  {\bibfield  {journal} {\bibinfo  {journal} {\protect\JournalTitle{Scientific
  reports}}\ }\textbf {\bibinfo {volume} {7}},\ \bibinfo {pages} {7643}
  (\bibinfo {year} {2017})}\BibitemShut {NoStop}%
\bibitem [{\citenamefont {Repula}\ \emph {et~al.}(2022)\citenamefont {Repula},
  \citenamefont {Abraham}, \citenamefont {Cherpak},\ and\ \citenamefont
  {Smalyukh}}]{repula2022}%
  \BibitemOpen
  \bibfield  {author} {\bibinfo {author} {\bibfnamefont {A.}~\bibnamefont
  {Repula}}, \bibinfo {author} {\bibfnamefont {E.}~\bibnamefont {Abraham}},
  \bibinfo {author} {\bibfnamefont {V.}~\bibnamefont {Cherpak}},\ and\ \bibinfo
  {author} {\bibfnamefont {I.~I.}\ \bibnamefont {Smalyukh}},\ }\bibfield
  {title} {\bibinfo {title} {\emph {Biotropic liquid crystal phase
  transformations in cellulose-producing bacterial communities}},\ }\href
  {https://doi.org/https://doi.org/10.1073/pnas.2200930119} {\bibfield
  {journal} {\bibinfo  {journal} {\protect\JournalTitle{Proceedings of the
  National Academy of Sciences}}\ }\textbf {\bibinfo {volume} {119}},\ \bibinfo
  {pages} {e2200930119} (\bibinfo {year} {2022})}\BibitemShut {NoStop}%
\bibitem [{\citenamefont {Gao}\ and\ \citenamefont {Wang}(2014)}]{gao2014}%
  \BibitemOpen
  \bibfield  {author} {\bibinfo {author} {\bibfnamefont {W.}~\bibnamefont
  {Gao}}\ and\ \bibinfo {author} {\bibfnamefont {J.}~\bibnamefont {Wang}},\
  }\bibfield  {title} {\bibinfo {title} {\emph {Synthetic micro/nanomotors in
  drug delivery}},\ }\href {https://doi.org/https://doi.org/10.1039/C4NR03124E}
  {\bibfield  {journal} {\bibinfo  {journal}
  {\protect\JournalTitle{Nanoscale}}\ }\textbf {\bibinfo {volume} {6}},\
  \bibinfo {pages} {10486} (\bibinfo {year} {2014})}\BibitemShut {NoStop}%
\bibitem [{\citenamefont {Rao}\ \emph {et~al.}(2015)\citenamefont {Rao},
  \citenamefont {Li}, \citenamefont {Meng}, \citenamefont {Zheng},
  \citenamefont {Cai},\ and\ \citenamefont {Wang}}]{rao2015}%
  \BibitemOpen
  \bibfield  {author} {\bibinfo {author} {\bibfnamefont {K.~J.}\ \bibnamefont
  {Rao}}, \bibinfo {author} {\bibfnamefont {F.}~\bibnamefont {Li}}, \bibinfo
  {author} {\bibfnamefont {L.}~\bibnamefont {Meng}}, \bibinfo {author}
  {\bibfnamefont {H.}~\bibnamefont {Zheng}}, \bibinfo {author} {\bibfnamefont
  {F.}~\bibnamefont {Cai}},\ and\ \bibinfo {author} {\bibfnamefont
  {W.}~\bibnamefont {Wang}},\ }\bibfield  {title} {\bibinfo {title} {\emph {A
  force to be reckoned with: a review of synthetic microswimmers powered by
  ultrasound}},\ }\href
  {https://doi.org/https://doi.org/10.1002/smll.201403621} {\bibfield
  {journal} {\bibinfo  {journal} {\protect\JournalTitle{Small}}\ }\textbf
  {\bibinfo {volume} {11}},\ \bibinfo {pages} {2836} (\bibinfo {year}
  {2015})}\BibitemShut {NoStop}%
\bibitem [{\citenamefont {Wu}\ \emph {et~al.}(2020)\citenamefont {Wu},
  \citenamefont {Chen}, \citenamefont {Mukasa}, \citenamefont {Pak},\ and\
  \citenamefont {Gao}}]{wu2020}%
  \BibitemOpen
  \bibfield  {author} {\bibinfo {author} {\bibfnamefont {Z.}~\bibnamefont
  {Wu}}, \bibinfo {author} {\bibfnamefont {Y.}~\bibnamefont {Chen}}, \bibinfo
  {author} {\bibfnamefont {D.}~\bibnamefont {Mukasa}}, \bibinfo {author}
  {\bibfnamefont {O.~S.}\ \bibnamefont {Pak}},\ and\ \bibinfo {author}
  {\bibfnamefont {W.}~\bibnamefont {Gao}},\ }\bibfield  {title} {\bibinfo
  {title} {\emph {Medical micro/nanorobots in complex media}},\ }\href
  {https://doi.org/https://doi.org/10.1039/D0CS00309C} {\bibfield  {journal}
  {\bibinfo  {journal} {\protect\JournalTitle{Chemical Society Reviews}}\
  }\textbf {\bibinfo {volume} {49}},\ \bibinfo {pages} {8088} (\bibinfo {year}
  {2020})}\BibitemShut {NoStop}%
\bibitem [{\citenamefont {Masocha}\ \emph {et~al.}(2004)\citenamefont
  {Masocha}, \citenamefont {Robertson}, \citenamefont {Rottenberg},
  \citenamefont {Mhlanga}, \citenamefont {Sorokin}, \citenamefont {Kristensson}
  \emph {et~al.}}]{masocha2004}%
  \BibitemOpen
  \bibfield  {author} {\bibinfo {author} {\bibfnamefont {W.}~\bibnamefont
  {Masocha}}, \bibinfo {author} {\bibfnamefont {B.}~\bibnamefont {Robertson}},
  \bibinfo {author} {\bibfnamefont {M.~E.}\ \bibnamefont {Rottenberg}},
  \bibinfo {author} {\bibfnamefont {J.}~\bibnamefont {Mhlanga}}, \bibinfo
  {author} {\bibfnamefont {L.}~\bibnamefont {Sorokin}}, \bibinfo {author}
  {\bibfnamefont {K.}~\bibnamefont {Kristensson}}, \emph {et~al.},\ }\bibfield
  {title} {\bibinfo {title} {\emph {Cerebral vessel laminins and IFN-$\gamma$
  define Trypanosoma brucei brucei penetration of the blood-brain barrier}},\
  }\href {https://doi.org/https://doi.org/10.1172/JCI22104} {\bibfield
  {journal} {\bibinfo  {journal} {\protect\JournalTitle{The Journal of clinical
  investigation}}\ }\textbf {\bibinfo {volume} {114}},\ \bibinfo {pages} {689}
  (\bibinfo {year} {2004})}\BibitemShut {NoStop}%
\bibitem [{\citenamefont {Patteson}\ \emph {et~al.}(2015)\citenamefont
  {Patteson}, \citenamefont {Gopinath}, \citenamefont {Goulian},\ and\
  \citenamefont {Arratia}}]{patteson2015}%
  \BibitemOpen
  \bibfield  {author} {\bibinfo {author} {\bibfnamefont {A.}~\bibnamefont
  {Patteson}}, \bibinfo {author} {\bibfnamefont {A.}~\bibnamefont {Gopinath}},
  \bibinfo {author} {\bibfnamefont {M.}~\bibnamefont {Goulian}},\ and\ \bibinfo
  {author} {\bibfnamefont {P.}~\bibnamefont {Arratia}},\ }\bibfield  {title}
  {\bibinfo {title} {\emph {Running and tumbling with E. coli in polymeric
  solutions}},\ }\href {https://doi.org/https://doi.org/10.1038/srep15761}
  {\bibfield  {journal} {\bibinfo  {journal} {\protect\JournalTitle{Scientific
  reports}}\ }\textbf {\bibinfo {volume} {5}},\ \bibinfo {pages} {15761}
  (\bibinfo {year} {2015})}\BibitemShut {NoStop}%
\bibitem [{\citenamefont {Gomez-Solano}\ \emph {et~al.}(2016)\citenamefont
  {Gomez-Solano}, \citenamefont {Blokhuis},\ and\ \citenamefont
  {Bechinger}}]{gomezsolano2016}%
  \BibitemOpen
  \bibfield  {author} {\bibinfo {author} {\bibfnamefont {J.~R.}\ \bibnamefont
  {Gomez-Solano}}, \bibinfo {author} {\bibfnamefont {A.}~\bibnamefont
  {Blokhuis}},\ and\ \bibinfo {author} {\bibfnamefont {C.}~\bibnamefont
  {Bechinger}},\ }\bibfield  {title} {\bibinfo {title} {\emph {Dynamics of
  Self-Propelled Janus Particles in Viscoelastic Fluids}},\ }\href
  {https://doi.org/10.1103/PhysRevLett.116.138301} {\bibfield  {journal}
  {\bibinfo  {journal} {\protect\JournalTitle{Phys. Rev. Lett.}}\ }\textbf
  {\bibinfo {volume} {116}},\ \bibinfo {pages} {138301} (\bibinfo {year}
  {2016})}\BibitemShut {NoStop}%
\bibitem [{\citenamefont {Tung}\ \emph {et~al.}(2017)\citenamefont {Tung},
  \citenamefont {Lin}, \citenamefont {Harvey}, \citenamefont {Fiore},
  \citenamefont {Ardon}, \citenamefont {Wu},\ and\ \citenamefont
  {Suarez}}]{tung2017}%
  \BibitemOpen
  \bibfield  {author} {\bibinfo {author} {\bibfnamefont {C.-k.}\ \bibnamefont
  {Tung}}, \bibinfo {author} {\bibfnamefont {C.}~\bibnamefont {Lin}}, \bibinfo
  {author} {\bibfnamefont {B.}~\bibnamefont {Harvey}}, \bibinfo {author}
  {\bibfnamefont {A.~G.}\ \bibnamefont {Fiore}}, \bibinfo {author}
  {\bibfnamefont {F.}~\bibnamefont {Ardon}}, \bibinfo {author} {\bibfnamefont
  {M.}~\bibnamefont {Wu}},\ and\ \bibinfo {author} {\bibfnamefont {S.~S.}\
  \bibnamefont {Suarez}},\ }\bibfield  {title} {\bibinfo {title} {\emph {Fluid
  viscoelasticity promotes collective swimming of sperm}},\ }\href
  {https://doi.org/https://doi.org/10.1038/s41598-017-03341-4} {\bibfield
  {journal} {\bibinfo  {journal} {\protect\JournalTitle{Scientific reports}}\
  }\textbf {\bibinfo {volume} {7}},\ \bibinfo {pages} {3152} (\bibinfo {year}
  {2017})}\BibitemShut {NoStop}%
\bibitem [{\citenamefont {Yuan}\ \emph {et~al.}(2018)\citenamefont {Yuan},
  \citenamefont {Zhao}, \citenamefont {Yan}, \citenamefont {Tang},
  \citenamefont {Alici}, \citenamefont {Zhang},\ and\ \citenamefont
  {Li}}]{yuan2018}%
  \BibitemOpen
  \bibfield  {author} {\bibinfo {author} {\bibfnamefont {D.}~\bibnamefont
  {Yuan}}, \bibinfo {author} {\bibfnamefont {Q.}~\bibnamefont {Zhao}}, \bibinfo
  {author} {\bibfnamefont {S.}~\bibnamefont {Yan}}, \bibinfo {author}
  {\bibfnamefont {S.-Y.}\ \bibnamefont {Tang}}, \bibinfo {author}
  {\bibfnamefont {G.}~\bibnamefont {Alici}}, \bibinfo {author} {\bibfnamefont
  {J.}~\bibnamefont {Zhang}},\ and\ \bibinfo {author} {\bibfnamefont
  {W.}~\bibnamefont {Li}},\ }\bibfield  {title} {\bibinfo {title} {\emph
  {Recent progress of particle migration in viscoelastic fluids}},\ }\href
  {https://doi.org/https://doi.org/10.1039/C7LC01076A} {\bibfield  {journal}
  {\bibinfo  {journal} {\protect\JournalTitle{Lab on a Chip}}\ }\textbf
  {\bibinfo {volume} {18}},\ \bibinfo {pages} {551} (\bibinfo {year}
  {2018})}\BibitemShut {NoStop}%
\bibitem [{\citenamefont {Ishimoto}\ and\ \citenamefont
  {Gaffney}(2018)}]{ishimoto2018}%
  \BibitemOpen
  \bibfield  {author} {\bibinfo {author} {\bibfnamefont {K.}~\bibnamefont
  {Ishimoto}}\ and\ \bibinfo {author} {\bibfnamefont {E.~A.}\ \bibnamefont
  {Gaffney}},\ }\bibfield  {title} {\bibinfo {title} {\emph {Hydrodynamic
  clustering of human sperm in viscoelastic fluids}},\ }\href
  {https://doi.org/https://doi.org/10.1038/s41598-018-33584-8} {\bibfield
  {journal} {\bibinfo  {journal} {\protect\JournalTitle{Scientific Reports}}\
  }\textbf {\bibinfo {volume} {8}},\ \bibinfo {pages} {15600} (\bibinfo {year}
  {2018})}\BibitemShut {NoStop}%
\bibitem [{\citenamefont {Liu}\ \emph {et~al.}(2021)\citenamefont {Liu},
  \citenamefont {Shankar}, \citenamefont {Marchetti},\ and\ \citenamefont
  {Wu}}]{liu2021Viscoelastic}%
  \BibitemOpen
  \bibfield  {author} {\bibinfo {author} {\bibfnamefont {S.}~\bibnamefont
  {Liu}}, \bibinfo {author} {\bibfnamefont {S.}~\bibnamefont {Shankar}},
  \bibinfo {author} {\bibfnamefont {M.~C.}\ \bibnamefont {Marchetti}},\ and\
  \bibinfo {author} {\bibfnamefont {Y.}~\bibnamefont {Wu}},\ }\bibfield
  {title} {\bibinfo {title} {\emph {Viscoelastic control of spatiotemporal
  order in bacterial active matter}},\ }\href
  {https://doi.org/https://doi.org/10.1038/s41586-020-03168-6} {\bibfield
  {journal} {\bibinfo  {journal} {\protect\JournalTitle{Nature}}\ }\textbf
  {\bibinfo {volume} {590}},\ \bibinfo {pages} {80} (\bibinfo {year}
  {2021})}\BibitemShut {NoStop}%
\bibitem [{\citenamefont {Peng}\ \emph {et~al.}(2016)\citenamefont {Peng},
  \citenamefont {Turiv}, \citenamefont {Guo}, \citenamefont {Wei},\ and\
  \citenamefont {Lavrentovich}}]{peng2016}%
  \BibitemOpen
  \bibfield  {author} {\bibinfo {author} {\bibfnamefont {C.}~\bibnamefont
  {Peng}}, \bibinfo {author} {\bibfnamefont {T.}~\bibnamefont {Turiv}},
  \bibinfo {author} {\bibfnamefont {Y.}~\bibnamefont {Guo}}, \bibinfo {author}
  {\bibfnamefont {Q.-H.}\ \bibnamefont {Wei}},\ and\ \bibinfo {author}
  {\bibfnamefont {O.~D.}\ \bibnamefont {Lavrentovich}},\ }\bibfield  {title}
  {\bibinfo {title} {\emph {Command of active matter by topological defects and
  patterns}},\ }\href {https://doi.org/10.1126/science.aah6936} {\bibfield
  {journal} {\bibinfo  {journal} {\protect\JournalTitle{Science}}\ }\textbf
  {\bibinfo {volume} {354}},\ \bibinfo {pages} {882} (\bibinfo {year}
  {2016})}\BibitemShut {NoStop}%
\bibitem [{\citenamefont {Kumar}\ \emph {et~al.}(2013)\citenamefont {Kumar},
  \citenamefont {Galstian}, \citenamefont {Pattanayek},\ and\ \citenamefont
  {Rainville}}]{kumar2013}%
  \BibitemOpen
  \bibfield  {author} {\bibinfo {author} {\bibfnamefont {A.}~\bibnamefont
  {Kumar}}, \bibinfo {author} {\bibfnamefont {T.}~\bibnamefont {Galstian}},
  \bibinfo {author} {\bibfnamefont {S.~K.}\ \bibnamefont {Pattanayek}},\ and\
  \bibinfo {author} {\bibfnamefont {S.}~\bibnamefont {Rainville}},\ }\bibfield
  {title} {\bibinfo {title} {\emph {The motility of bacteria in an anisotropic
  liquid environment}},\ }\href
  {https://doi.org/https://doi.org/10.1080/15421406.2012.762493} {\bibfield
  {journal} {\bibinfo  {journal} {\protect\JournalTitle{Molecular Crystals and
  Liquid Crystals}}\ }\textbf {\bibinfo {volume} {574}},\ \bibinfo {pages} {33}
  (\bibinfo {year} {2013})}\BibitemShut {NoStop}%
\bibitem [{\citenamefont {Zhou}\ \emph {et~al.}(2014)\citenamefont {Zhou},
  \citenamefont {Sokolov}, \citenamefont {Lavrentovich},\ and\ \citenamefont
  {Aranson}}]{zhou2014}%
  \BibitemOpen
  \bibfield  {author} {\bibinfo {author} {\bibfnamefont {S.}~\bibnamefont
  {Zhou}}, \bibinfo {author} {\bibfnamefont {A.}~\bibnamefont {Sokolov}},
  \bibinfo {author} {\bibfnamefont {O.~D.}\ \bibnamefont {Lavrentovich}},\ and\
  \bibinfo {author} {\bibfnamefont {I.~S.}\ \bibnamefont {Aranson}},\
  }\bibfield  {title} {\bibinfo {title} {\emph {Living liquid crystals}},\
  }\href {https://doi.org/10.1073/pnas.1321926111} {\bibfield  {journal}
  {\bibinfo  {journal} {\protect\JournalTitle{Proceedings of the National
  Academy of Sciences}}\ }\textbf {\bibinfo {volume} {111}},\ \bibinfo {pages}
  {1265} (\bibinfo {year} {2014})}\BibitemShut {NoStop}%
\bibitem [{\citenamefont {Sokolov}\ \emph {et~al.}(2015)\citenamefont
  {Sokolov}, \citenamefont {Zhou}, \citenamefont {Lavrentovich},\ and\
  \citenamefont {Aranson}}]{sokolov2015}%
  \BibitemOpen
  \bibfield  {author} {\bibinfo {author} {\bibfnamefont {A.}~\bibnamefont
  {Sokolov}}, \bibinfo {author} {\bibfnamefont {S.}~\bibnamefont {Zhou}},
  \bibinfo {author} {\bibfnamefont {O.~D.}\ \bibnamefont {Lavrentovich}},\ and\
  \bibinfo {author} {\bibfnamefont {I.~S.}\ \bibnamefont {Aranson}},\
  }\bibfield  {title} {\bibinfo {title} {\emph {Individual behavior and
  pairwise interactions between microswimmers in anisotropic liquid}},\ }\href
  {https://doi.org/10.1103/PhysRevE.91.013009} {\bibfield  {journal} {\bibinfo
  {journal} {\protect\JournalTitle{Phys. Rev. E}}\ }\textbf {\bibinfo {volume}
  {91}},\ \bibinfo {pages} {013009} (\bibinfo {year} {2015})}\BibitemShut
  {NoStop}%
\bibitem [{\citenamefont {Koizumi}\ \emph {et~al.}(2020)\citenamefont
  {Koizumi}, \citenamefont {Turiv}, \citenamefont {Genkin}, \citenamefont
  {Lastowski}, \citenamefont {Yu}, \citenamefont {Chaganava}, \citenamefont
  {Wei}, \citenamefont {Aranson},\ and\ \citenamefont
  {Lavrentovich}}]{koizumi2020}%
  \BibitemOpen
  \bibfield  {author} {\bibinfo {author} {\bibfnamefont {R.}~\bibnamefont
  {Koizumi}}, \bibinfo {author} {\bibfnamefont {T.}~\bibnamefont {Turiv}},
  \bibinfo {author} {\bibfnamefont {M.~M.}\ \bibnamefont {Genkin}}, \bibinfo
  {author} {\bibfnamefont {R.~J.}\ \bibnamefont {Lastowski}}, \bibinfo {author}
  {\bibfnamefont {H.}~\bibnamefont {Yu}}, \bibinfo {author} {\bibfnamefont
  {I.}~\bibnamefont {Chaganava}}, \bibinfo {author} {\bibfnamefont {Q.-H.}\
  \bibnamefont {Wei}}, \bibinfo {author} {\bibfnamefont {I.~S.}\ \bibnamefont
  {Aranson}},\ and\ \bibinfo {author} {\bibfnamefont {O.~D.}\ \bibnamefont
  {Lavrentovich}},\ }\bibfield  {title} {\bibinfo {title} {\emph {Control of
  microswimmers by spiral nematic vortices: Transition from individual to
  collective motion and contraction, expansion, and stable circulation of
  bacterial swirls}},\ }\href
  {https://doi.org/10.1103/PhysRevResearch.2.033060} {\bibfield  {journal}
  {\bibinfo  {journal} {\protect\JournalTitle{Phys. Rev. Res.}}\ }\textbf
  {\bibinfo {volume} {2}},\ \bibinfo {pages} {033060} (\bibinfo {year}
  {2020})}\BibitemShut {NoStop}%
\bibitem [{\citenamefont {Genkin}\ \emph {et~al.}(2017)\citenamefont {Genkin},
  \citenamefont {Sokolov}, \citenamefont {Lavrentovich},\ and\ \citenamefont
  {Aranson}}]{genkin2017}%
  \BibitemOpen
  \bibfield  {author} {\bibinfo {author} {\bibfnamefont {M.~M.}\ \bibnamefont
  {Genkin}}, \bibinfo {author} {\bibfnamefont {A.}~\bibnamefont {Sokolov}},
  \bibinfo {author} {\bibfnamefont {O.~D.}\ \bibnamefont {Lavrentovich}},\ and\
  \bibinfo {author} {\bibfnamefont {I.~S.}\ \bibnamefont {Aranson}},\
  }\bibfield  {title} {\bibinfo {title} {\emph {Topological Defects in a Living
  Nematic Ensnare Swimming Bacteria}},\ }\href
  {https://doi.org/10.1103/PhysRevX.7.011029} {\bibfield  {journal} {\bibinfo
  {journal} {\protect\JournalTitle{Phys. Rev. X}}\ }\textbf {\bibinfo {volume}
  {7}},\ \bibinfo {pages} {011029} (\bibinfo {year} {2017})}\BibitemShut
  {NoStop}%
\bibitem [{\citenamefont {Genkin}\ \emph {et~al.}(2018)\citenamefont {Genkin},
  \citenamefont {Sokolov},\ and\ \citenamefont {Aranson}}]{genkin2018}%
  \BibitemOpen
  \bibfield  {author} {\bibinfo {author} {\bibfnamefont {M.~M.}\ \bibnamefont
  {Genkin}}, \bibinfo {author} {\bibfnamefont {A.}~\bibnamefont {Sokolov}},\
  and\ \bibinfo {author} {\bibfnamefont {I.~S.}\ \bibnamefont {Aranson}},\
  }\bibfield  {title} {\bibinfo {title} {\emph {Spontaneous topological
  charging of tactoids in a living nematic}},\ }\href
  {https://doi.org/10.1088/1367-2630/aab1a3} {\bibfield  {journal} {\bibinfo
  {journal} {\protect\JournalTitle{New J. Phys.}}\ }\textbf {\bibinfo {volume}
  {20}},\ \bibinfo {pages} {43027} (\bibinfo {year} {2018})}\BibitemShut
  {NoStop}%
\bibitem [{\citenamefont {Turiv}\ \emph {et~al.}(2020)\citenamefont {Turiv},
  \citenamefont {Koizumi}, \citenamefont {Thijssen}, \citenamefont {Genkin},
  \citenamefont {Yu}, \citenamefont {Peng}, \citenamefont {Wei}, \citenamefont
  {Yeomans}, \citenamefont {Aranson}, \citenamefont {Doostmohammadi},\ and\
  \citenamefont {Lavrentovich}}]{turiv2020}%
  \BibitemOpen
  \bibfield  {author} {\bibinfo {author} {\bibfnamefont {T.}~\bibnamefont
  {Turiv}}, \bibinfo {author} {\bibfnamefont {R.}~\bibnamefont {Koizumi}},
  \bibinfo {author} {\bibfnamefont {K.}~\bibnamefont {Thijssen}}, \bibinfo
  {author} {\bibfnamefont {M.~M.}\ \bibnamefont {Genkin}}, \bibinfo {author}
  {\bibfnamefont {H.}~\bibnamefont {Yu}}, \bibinfo {author} {\bibfnamefont
  {C.}~\bibnamefont {Peng}}, \bibinfo {author} {\bibfnamefont {Q.-H.}\
  \bibnamefont {Wei}}, \bibinfo {author} {\bibfnamefont {J.~M.}\ \bibnamefont
  {Yeomans}}, \bibinfo {author} {\bibfnamefont {I.~S.}\ \bibnamefont
  {Aranson}}, \bibinfo {author} {\bibfnamefont {A.}~\bibnamefont
  {Doostmohammadi}},\ and\ \bibinfo {author} {\bibfnamefont {O.~D.}\
  \bibnamefont {Lavrentovich}},\ }\bibfield  {title} {\bibinfo {title} {\emph
  {Polar jets of swimming bacteria condensed by a patterned liquid crystal}},\
  }\href {https://doi.org/10.1038/s41567-020-0793-0} {\bibfield  {journal}
  {\bibinfo  {journal} {\protect\JournalTitle{Nat. Phys.}}\ }\textbf {\bibinfo
  {volume} {16}},\ \bibinfo {pages} {481} (\bibinfo {year} {2020})}\BibitemShut
  {NoStop}%
\bibitem [{\citenamefont {Mushenheim}\ \emph {et~al.}(2013)\citenamefont
  {Mushenheim}, \citenamefont {Trivedi}, \citenamefont {Tuson}, \citenamefont
  {Weibel},\ and\ \citenamefont {Abbott}}]{mushenheim2013}%
  \BibitemOpen
  \bibfield  {author} {\bibinfo {author} {\bibfnamefont {P.~C.}\ \bibnamefont
  {Mushenheim}}, \bibinfo {author} {\bibfnamefont {R.~R.}\ \bibnamefont
  {Trivedi}}, \bibinfo {author} {\bibfnamefont {H.~H.}\ \bibnamefont {Tuson}},
  \bibinfo {author} {\bibfnamefont {D.~B.}\ \bibnamefont {Weibel}},\ and\
  \bibinfo {author} {\bibfnamefont {N.~L.}\ \bibnamefont {Abbott}},\ }\bibfield
   {title} {\bibinfo {title} {\emph {Dynamic self-assembly of motile bacteria
  in liquid crystals}},\ }\href {https://doi.org/10.1039/C3SM52423J} {\bibfield
   {journal} {\bibinfo  {journal} {\protect\JournalTitle{Soft Matter}}\
  }\textbf {\bibinfo {volume} {10}},\ \bibinfo {pages} {88} (\bibinfo {year}
  {2013})}\BibitemShut {NoStop}%
\bibitem [{\citenamefont {Mushenheim}\ \emph {et~al.}(2015)\citenamefont
  {Mushenheim}, \citenamefont {Trivedi}, \citenamefont {Roy}, \citenamefont
  {Arnold}, \citenamefont {Weibel},\ and\ \citenamefont
  {Abbott}}]{mushenheim2015}%
  \BibitemOpen
  \bibfield  {author} {\bibinfo {author} {\bibfnamefont {P.~C.}\ \bibnamefont
  {Mushenheim}}, \bibinfo {author} {\bibfnamefont {R.~R.}\ \bibnamefont
  {Trivedi}}, \bibinfo {author} {\bibfnamefont {S.~S.}\ \bibnamefont {Roy}},
  \bibinfo {author} {\bibfnamefont {M.~S.}\ \bibnamefont {Arnold}}, \bibinfo
  {author} {\bibfnamefont {D.~B.}\ \bibnamefont {Weibel}},\ and\ \bibinfo
  {author} {\bibfnamefont {N.~L.}\ \bibnamefont {Abbott}},\ }\bibfield  {title}
  {\bibinfo {title} {\emph {Effects of confinement, surface-induced
  orientations and strain on dynamical behaviors of bacteria in thin liquid
  crystalline films}},\ }\href
  {https://pubs.rsc.org/en/content/articlelanding/2015/sm/c5sm01489a}
  {\bibfield  {journal} {\bibinfo  {journal} {\protect\JournalTitle{Soft
  Matter}}\ }\textbf {\bibinfo {volume} {11}},\ \bibinfo {pages} {6821}
  (\bibinfo {year} {2015})}\BibitemShut {NoStop}%
\bibitem [{\citenamefont {Trivedi}\ \emph {et~al.}(2015)\citenamefont
  {Trivedi}, \citenamefont {Maeda}, \citenamefont {Abbott}, \citenamefont
  {Spagnolie},\ and\ \citenamefont {Weibel}}]{trivedi2015}%
  \BibitemOpen
  \bibfield  {author} {\bibinfo {author} {\bibfnamefont {R.~R.}\ \bibnamefont
  {Trivedi}}, \bibinfo {author} {\bibfnamefont {R.}~\bibnamefont {Maeda}},
  \bibinfo {author} {\bibfnamefont {N.~L.}\ \bibnamefont {Abbott}}, \bibinfo
  {author} {\bibfnamefont {S.~E.}\ \bibnamefont {Spagnolie}},\ and\ \bibinfo
  {author} {\bibfnamefont {D.~B.}\ \bibnamefont {Weibel}},\ }\bibfield  {title}
  {\bibinfo {title} {\emph {Bacterial transport of colloids in liquid
  crystalline environments}},\ }\href
  {https://pubs.rsc.org/en/content/articlelanding/2015/sm/c5sm02041g}
  {\bibfield  {journal} {\bibinfo  {journal} {\protect\JournalTitle{Soft
  Matter}}\ }\textbf {\bibinfo {volume} {11}},\ \bibinfo {pages} {8404}
  (\bibinfo {year} {2015})}\BibitemShut {NoStop}%
\bibitem [{\citenamefont {Lintuvuori}\ \emph {et~al.}(2017)\citenamefont
  {Lintuvuori}, \citenamefont {Würger},\ and\ \citenamefont
  {Stratford}}]{lintuvuori2017}%
  \BibitemOpen
  \bibfield  {author} {\bibinfo {author} {\bibfnamefont {J.}~\bibnamefont
  {Lintuvuori}}, \bibinfo {author} {\bibfnamefont {A.}~\bibnamefont
  {Würger}},\ and\ \bibinfo {author} {\bibfnamefont {K.}~\bibnamefont
  {Stratford}},\ }\bibfield  {title} {\bibinfo {title} {\emph {Hydrodynamics
  Defines the Stable Swimming Direction of Spherical Squirmers in a Nematic
  Liquid Crystal}},\ }\href {https://doi.org/10.1103/PhysRevLett.119.068001}
  {\bibfield  {journal} {\bibinfo  {journal} {\protect\JournalTitle{Phys. Rev.
  Lett.}}\ }\textbf {\bibinfo {volume} {119}},\ \bibinfo {pages} {68001}
  (\bibinfo {year} {2017})},\ \bibinfo {note} {publisher: American Physical
  Society}\BibitemShut {NoStop}%
\bibitem [{\citenamefont {Gautam}\ and\ \citenamefont
  {Lintuvuori}(2024)}]{gautam2024}%
  \BibitemOpen
  \bibfield  {author} {\bibinfo {author} {\bibfnamefont {B.}~\bibnamefont
  {Gautam}}\ and\ \bibinfo {author} {\bibfnamefont {J.~S.}\ \bibnamefont
  {Lintuvuori}},\ }\bibfield  {title} {\bibinfo {title} {\emph {Microswimmers
  Knead Nematics into Cholesterics}},\ }\href
  {https://doi.org/10.1103/PhysRevLett.132.238301} {\bibfield  {journal}
  {\bibinfo  {journal} {\protect\JournalTitle{Phys. Rev. Lett.}}\ }\textbf
  {\bibinfo {volume} {132}},\ \bibinfo {pages} {238301} (\bibinfo {year}
  {2024})}\BibitemShut {NoStop}%
\bibitem [{\citenamefont {Chi}\ \emph {et~al.}(2020)\citenamefont {Chi},
  \citenamefont {Potomkin}, \citenamefont {Zhang}, \citenamefont {Berlyand},\
  and\ \citenamefont {Aranson}}]{chi2020}%
  \BibitemOpen
  \bibfield  {author} {\bibinfo {author} {\bibfnamefont {H.}~\bibnamefont
  {Chi}}, \bibinfo {author} {\bibfnamefont {M.}~\bibnamefont {Potomkin}},
  \bibinfo {author} {\bibfnamefont {L.}~\bibnamefont {Zhang}}, \bibinfo
  {author} {\bibfnamefont {L.}~\bibnamefont {Berlyand}},\ and\ \bibinfo
  {author} {\bibfnamefont {I.~S.}\ \bibnamefont {Aranson}},\ }\bibfield
  {title} {\bibinfo {title} {\emph {Surface anchoring controls orientation of a
  microswimmer in nematic liquid crystal}},\ }\href
  {https://doi.org/https://doi.org/10.1038/s42005-020-00432-z} {\bibfield
  {journal} {\bibinfo  {journal} {\protect\JournalTitle{Communications
  Physics}}\ }\textbf {\bibinfo {volume} {3}},\ \bibinfo {pages} {162}
  (\bibinfo {year} {2020})}\BibitemShut {NoStop}%
\bibitem [{\citenamefont {Zhou}\ \emph {et~al.}(2017)\citenamefont {Zhou},
  \citenamefont {Tovkach}, \citenamefont {Golovaty}, \citenamefont {Sokolov},
  \citenamefont {Aranson},\ and\ \citenamefont {Lavrentovich}}]{zhou2017}%
  \BibitemOpen
  \bibfield  {author} {\bibinfo {author} {\bibfnamefont {S.}~\bibnamefont
  {Zhou}}, \bibinfo {author} {\bibfnamefont {O.}~\bibnamefont {Tovkach}},
  \bibinfo {author} {\bibfnamefont {D.}~\bibnamefont {Golovaty}}, \bibinfo
  {author} {\bibfnamefont {A.}~\bibnamefont {Sokolov}}, \bibinfo {author}
  {\bibfnamefont {I.~S.}\ \bibnamefont {Aranson}},\ and\ \bibinfo {author}
  {\bibfnamefont {O.~D.}\ \bibnamefont {Lavrentovich}},\ }\bibfield  {title}
  {\bibinfo {title} {\emph {Dynamic states of swimming bacteria in a nematic
  liquid crystal cell with homeotropic alignment}},\ }\href
  {https://doi.org/https://doi.org/10.1088/1367-2630/aa695b} {\bibfield
  {journal} {\bibinfo  {journal} {\protect\JournalTitle{New Journal of
  Physics}}\ }\textbf {\bibinfo {volume} {19}},\ \bibinfo {pages} {055006}
  (\bibinfo {year} {2017})}\BibitemShut {NoStop}%
\bibitem [{\citenamefont {Daddi-Moussa-Ider}\ and\ \citenamefont
  {Menzel}(2018)}]{daddi2018}%
  \BibitemOpen
  \bibfield  {author} {\bibinfo {author} {\bibfnamefont {A.}~\bibnamefont
  {Daddi-Moussa-Ider}}\ and\ \bibinfo {author} {\bibfnamefont {A.~M.}\
  \bibnamefont {Menzel}},\ }\bibfield  {title} {\bibinfo {title} {\emph
  {Dynamics of a simple model microswimmer in an anisotropic fluid:
  Implications for alignment behavior and active transport in a nematic liquid
  crystal}},\ }\href
  {https://doi.org/https://doi.org/10.1103/PhysRevFluids.3.094102} {\bibfield
  {journal} {\bibinfo  {journal} {\protect\JournalTitle{Physical Review
  Fluids}}\ }\textbf {\bibinfo {volume} {3}},\ \bibinfo {pages} {094102}
  (\bibinfo {year} {2018})}\BibitemShut {NoStop}%
\bibitem [{\citenamefont {Ryan}\ \emph {et~al.}(2013)\citenamefont {Ryan},
  \citenamefont {Berlyand}, \citenamefont {Haines},\ and\ \citenamefont
  {Karpeev}}]{ryan2013}%
  \BibitemOpen
  \bibfield  {author} {\bibinfo {author} {\bibfnamefont {S.~D.}\ \bibnamefont
  {Ryan}}, \bibinfo {author} {\bibfnamefont {L.}~\bibnamefont {Berlyand}},
  \bibinfo {author} {\bibfnamefont {B.~M.}\ \bibnamefont {Haines}},\ and\
  \bibinfo {author} {\bibfnamefont {D.~A.}\ \bibnamefont {Karpeev}},\
  }\bibfield  {title} {\bibinfo {title} {\emph {A Kinetic Model for Semidilute
  Bacterial Suspensions}},\ }\href {https://doi.org/10.1137/120900575}
  {\bibfield  {journal} {\bibinfo  {journal} {\protect\JournalTitle{Multiscale
  Model. Simul.}}\ }\textbf {\bibinfo {volume} {11}},\ \bibinfo {pages} {1176}
  (\bibinfo {year} {2013})}\BibitemShut {NoStop}%
\bibitem [{\citenamefont {Lee}\ and\ \citenamefont
  {Mazza}(2015)}]{lee2015stochastic}%
  \BibitemOpen
  \bibfield  {author} {\bibinfo {author} {\bibfnamefont {K.-W.}\ \bibnamefont
  {Lee}}\ and\ \bibinfo {author} {\bibfnamefont {M.~G.}\ \bibnamefont
  {Mazza}},\ }\bibfield  {title} {\bibinfo {title} {\emph {Stochastic rotation
  dynamics for nematic liquid crystals}},\ }\bibfield  {journal} {\bibinfo
  {journal} {\protect\JournalTitle{The Journal of chemical physics}}\ }\textbf
  {\bibinfo {volume} {142}},\ \href
  {https://doi.org/https://doi.org/10.1063/1.4919310}
  {https://doi.org/10.1063/1.4919310} (\bibinfo {year} {2015})\BibitemShut
  {NoStop}%
\bibitem [{\citenamefont {Shendruk}\ and\ \citenamefont
  {Yeomans}(2015)}]{shendruk2015multi}%
  \BibitemOpen
  \bibfield  {author} {\bibinfo {author} {\bibfnamefont {T.~N.}\ \bibnamefont
  {Shendruk}}\ and\ \bibinfo {author} {\bibfnamefont {J.~M.}\ \bibnamefont
  {Yeomans}},\ }\bibfield  {title} {\bibinfo {title} {\emph {Multi-particle
  collision dynamics algorithm for nematic fluids}},\ }\href
  {https://doi.org/https://doi.org/10.1039/C5SM00839E} {\bibfield  {journal}
  {\bibinfo  {journal} {\protect\JournalTitle{Soft Matter}}\ }\textbf {\bibinfo
  {volume} {11}},\ \bibinfo {pages} {5101} (\bibinfo {year}
  {2015})}\BibitemShut {NoStop}%
\bibitem [{\citenamefont {Mandal}\ and\ \citenamefont
  {Mazza}(2019)}]{mandal2019multiparticle}%
  \BibitemOpen
  \bibfield  {author} {\bibinfo {author} {\bibfnamefont {S.}~\bibnamefont
  {Mandal}}\ and\ \bibinfo {author} {\bibfnamefont {M.~G.}\ \bibnamefont
  {Mazza}},\ }\bibfield  {title} {\bibinfo {title} {\emph {Multiparticle
  collision dynamics for tensorial nematodynamics}},\ }\href
  {https://doi.org/https://doi.org/10.1103/PhysRevE.99.063319} {\bibfield
  {journal} {\bibinfo  {journal} {\protect\JournalTitle{Physical Review E}}\
  }\textbf {\bibinfo {volume} {99}},\ \bibinfo {pages} {063319} (\bibinfo
  {year} {2019})}\BibitemShut {NoStop}%
\bibitem [{\citenamefont {H{\'\i}jar}(2020)}]{hijar2020dynamics}%
  \BibitemOpen
  \bibfield  {author} {\bibinfo {author} {\bibfnamefont {H.}~\bibnamefont
  {H{\'\i}jar}},\ }\bibfield  {title} {\bibinfo {title} {\emph {Dynamics of
  defects around anisotropic particles in nematic liquid crystals under
  shear}},\ }\href
  {https://doi.org/https://doi.org/10.1103/PhysRevE.102.062705} {\bibfield
  {journal} {\bibinfo  {journal} {\protect\JournalTitle{Physical Review E}}\
  }\textbf {\bibinfo {volume} {102}},\ \bibinfo {pages} {062705} (\bibinfo
  {year} {2020})}\BibitemShut {NoStop}%
\bibitem [{\citenamefont {Mandal}\ and\ \citenamefont
  {Mazza}(2021)}]{mandal2021}%
  \BibitemOpen
  \bibfield  {author} {\bibinfo {author} {\bibfnamefont {S.}~\bibnamefont
  {Mandal}}\ and\ \bibinfo {author} {\bibfnamefont {M.~G.}\ \bibnamefont
  {Mazza}},\ }\bibfield  {title} {\bibinfo {title} {\emph {Multiparticle
  collision dynamics simulations of a squirmer in a nematic fluid}},\ }\href
  {https://doi.org/https://doi.org/10.1140/epje/s10189-021-00072-3} {\bibfield
  {journal} {\bibinfo  {journal} {\protect\JournalTitle{The European Physical
  Journal E}}\ }\textbf {\bibinfo {volume} {44}},\ \bibinfo {pages} {64}
  (\bibinfo {year} {2021})}\BibitemShut {NoStop}%
\bibitem [{\citenamefont {Kozhukhov}\ and\ \citenamefont
  {Shendruk}(2022)}]{kozhukhov2022}%
  \BibitemOpen
  \bibfield  {author} {\bibinfo {author} {\bibfnamefont {T.}~\bibnamefont
  {Kozhukhov}}\ and\ \bibinfo {author} {\bibfnamefont {T.~N.}\ \bibnamefont
  {Shendruk}},\ }\bibfield  {title} {\bibinfo {title} {\emph {Mesoscopic
  simulations of active nematics}},\ }\href
  {https://doi.org/https://doi.org/10.1126/sciadv.abo5788} {\bibfield
  {journal} {\bibinfo  {journal} {\protect\JournalTitle{Science Advances}}\
  }\textbf {\bibinfo {volume} {8}},\ \bibinfo {pages} {eabo5788} (\bibinfo
  {year} {2022})}\BibitemShut {NoStop}%
\bibitem [{\citenamefont {Mac{\'\i}as-Dur{\'a}n}\ \emph
  {et~al.}(2023)\citenamefont {Mac{\'\i}as-Dur{\'a}n}, \citenamefont
  {Duarte-Alaniz},\ and\ \citenamefont {H{\'\i}jar}}]{macias2023}%
  \BibitemOpen
  \bibfield  {author} {\bibinfo {author} {\bibfnamefont {J.}~\bibnamefont
  {Mac{\'\i}as-Dur{\'a}n}}, \bibinfo {author} {\bibfnamefont {V.}~\bibnamefont
  {Duarte-Alaniz}},\ and\ \bibinfo {author} {\bibfnamefont {H.}~\bibnamefont
  {H{\'\i}jar}},\ }\bibfield  {title} {\bibinfo {title} {\emph {Active nematic
  liquid crystals simulated by particle-based mesoscopic methods}},\ }\href
  {https://doi.org/https://doi.org/10.1039/D3SM00481C} {\bibfield  {journal}
  {\bibinfo  {journal} {\protect\JournalTitle{Soft Matter}}\ }\textbf {\bibinfo
  {volume} {19}},\ \bibinfo {pages} {8052} (\bibinfo {year}
  {2023})}\BibitemShut {NoStop}%
\bibitem [{\citenamefont {Kozhukhov}\ \emph {et~al.}(2024)\citenamefont
  {Kozhukhov}, \citenamefont {Loewe},\ and\ \citenamefont
  {Shendruk}}]{kozhukhov2024}%
  \BibitemOpen
  \bibfield  {author} {\bibinfo {author} {\bibfnamefont {T.}~\bibnamefont
  {Kozhukhov}}, \bibinfo {author} {\bibfnamefont {B.}~\bibnamefont {Loewe}},\
  and\ \bibinfo {author} {\bibfnamefont {T.~N.}\ \bibnamefont {Shendruk}},\
  }\bibfield  {title} {\bibinfo {title} {\emph {Mitigating density fluctuations
  in particle-based active nematic simulations}},\ }\href@noop {} {\bibfield
  {journal} {\bibinfo  {journal} {\protect\JournalTitle{Communications
  Physics}}\ }\textbf {\bibinfo {volume} {7}},\ \bibinfo {pages} {251}
  (\bibinfo {year} {2024})}\BibitemShut {NoStop}%
\bibitem [{\citenamefont {Mandal}\ \emph {et~al.}(2025)\citenamefont {Mandal},
  \citenamefont {Mason}, \citenamefont {Croft},\ and\ \citenamefont
  {Mazza}}]{mandal2025}%
  \BibitemOpen
  \bibfield  {author} {\bibinfo {author} {\bibfnamefont {S.}~\bibnamefont
  {Mandal}}, \bibinfo {author} {\bibfnamefont {T.~J.}\ \bibnamefont {Mason}},
  \bibinfo {author} {\bibfnamefont {A.~C.}\ \bibnamefont {Croft}},\ and\
  \bibinfo {author} {\bibfnamefont {M.~G.}\ \bibnamefont {Mazza}},\ }\bibfield
  {title} {\bibinfo {title} {\emph {Cooperativity of Confined Nematic
  Microswimmers: From One to Many}},\ }\href
  {https://doi.org/https://doi.org/10.1103/PhysRevLett.134.128302} {\bibfield
  {journal} {\bibinfo  {journal} {\protect\JournalTitle{Physical Review
  Letters}}\ }\textbf {\bibinfo {volume} {134}},\ \bibinfo {pages} {128302}
  (\bibinfo {year} {2025})}\BibitemShut {NoStop}%
\bibitem [{\citenamefont {Denniston}\ \emph {et~al.}(2001)\citenamefont
  {Denniston}, \citenamefont {Orlandini},\ and\ \citenamefont
  {Yeomans}}]{denniston2001}%
  \BibitemOpen
  \bibfield  {author} {\bibinfo {author} {\bibfnamefont {C.}~\bibnamefont
  {Denniston}}, \bibinfo {author} {\bibfnamefont {E.}~\bibnamefont
  {Orlandini}},\ and\ \bibinfo {author} {\bibfnamefont {J.}~\bibnamefont
  {Yeomans}},\ }\bibfield  {title} {\bibinfo {title} {\emph {Lattice Boltzmann
  simulations of liquid crystal hydrodynamics}},\ }\href
  {https://doi.org/10.1103/PhysRevE.63.056702} {\bibfield  {journal} {\bibinfo
  {journal} {\protect\JournalTitle{Physical Review E}}\ }\textbf {\bibinfo
  {volume} {63}},\ \bibinfo {pages} {056702} (\bibinfo {year}
  {2001})}\BibitemShut {NoStop}%
\bibitem [{\citenamefont {Li}\ \emph {et~al.}(2019)\citenamefont {Li},
  \citenamefont {Shi}, \citenamefont {Huang}, \citenamefont {Chen},
  \citenamefont {Xiao}, \citenamefont {Liu}, \citenamefont {Chat{\'e}},\ and\
  \citenamefont {Zhang}}]{li2019}%
  \BibitemOpen
  \bibfield  {author} {\bibinfo {author} {\bibfnamefont {H.}~\bibnamefont
  {Li}}, \bibinfo {author} {\bibfnamefont {X.-q.}\ \bibnamefont {Shi}},
  \bibinfo {author} {\bibfnamefont {M.}~\bibnamefont {Huang}}, \bibinfo
  {author} {\bibfnamefont {X.}~\bibnamefont {Chen}}, \bibinfo {author}
  {\bibfnamefont {M.}~\bibnamefont {Xiao}}, \bibinfo {author} {\bibfnamefont
  {C.}~\bibnamefont {Liu}}, \bibinfo {author} {\bibfnamefont {H.}~\bibnamefont
  {Chat{\'e}}},\ and\ \bibinfo {author} {\bibfnamefont {H.}~\bibnamefont
  {Zhang}},\ }\bibfield  {title} {\bibinfo {title} {\emph {Data-driven
  quantitative modeling of bacterial active nematics}},\ }\href
  {https://doi.org/https://doi.org/10.1073/pnas.1812570116} {\bibfield
  {journal} {\bibinfo  {journal} {\protect\JournalTitle{Proceedings of the
  National Academy of Sciences}}\ }\textbf {\bibinfo {volume} {116}},\ \bibinfo
  {pages} {777} (\bibinfo {year} {2019})}\BibitemShut {NoStop}%
\bibitem [{\citenamefont {Li}\ \emph {et~al.}(2024)\citenamefont {Li},
  \citenamefont {Chat\'e}, \citenamefont {Sano}, \citenamefont {Shi},\ and\
  \citenamefont {Zhang}}]{li2024}%
  \BibitemOpen
  \bibfield  {author} {\bibinfo {author} {\bibfnamefont {H.}~\bibnamefont
  {Li}}, \bibinfo {author} {\bibfnamefont {H.}~\bibnamefont {Chat\'e}},
  \bibinfo {author} {\bibfnamefont {M.}~\bibnamefont {Sano}}, \bibinfo {author}
  {\bibfnamefont {X.-q.}\ \bibnamefont {Shi}},\ and\ \bibinfo {author}
  {\bibfnamefont {H.~P.}\ \bibnamefont {Zhang}},\ }\bibfield  {title} {\bibinfo
  {title} {\emph {Robust Edge Flows in Swarming Bacterial Colonies}},\ }\href
  {https://doi.org/10.1103/PhysRevX.14.041006} {\bibfield  {journal} {\bibinfo
  {journal} {\protect\JournalTitle{Phys. Rev. X}}\ }\textbf {\bibinfo {volume}
  {14}},\ \bibinfo {pages} {041006} (\bibinfo {year} {2024})}\BibitemShut
  {NoStop}%
\bibitem [{\citenamefont {Dai}(2015)}]{dai2015}%
  \BibitemOpen
  \bibfield  {author} {\bibinfo {author} {\bibfnamefont {J.~S.}\ \bibnamefont
  {Dai}},\ }\bibfield  {title} {\bibinfo {title} {\emph {Euler--Rodrigues
  formula variations, quaternion conjugation and intrinsic connections}},\
  }\href {https://doi.org/https://doi.org/10.1016/j.mechmachtheory.2015.03.004}
  {\bibfield  {journal} {\bibinfo  {journal} {\protect\JournalTitle{Mechanism
  and Machine Theory}}\ }\textbf {\bibinfo {volume} {92}},\ \bibinfo {pages}
  {144} (\bibinfo {year} {2015})}\BibitemShut {NoStop}%
\bibitem [{\citenamefont {Ishimoto}(2023)}]{ishimoto2023}%
  \BibitemOpen
  \bibfield  {author} {\bibinfo {author} {\bibfnamefont {K.}~\bibnamefont
  {Ishimoto}},\ }\bibfield  {title} {\bibinfo {title} {\emph {Jeffery’s
  orbits and microswimmers in flows: A theoretical review}},\ }\href
  {https://journals.jps.jp/doi/10.7566/JPSJ.92.062001} {\bibfield  {journal}
  {\bibinfo  {journal} {\protect\JournalTitle{Journal of the Physical Society
  of Japan}}\ }\textbf {\bibinfo {volume} {92}},\ \bibinfo {pages} {062001}
  (\bibinfo {year} {2023})}\BibitemShut {NoStop}%
\bibitem [{\citenamefont {Zhang}\ \emph {et~al.}(2016)\citenamefont {Zhang},
  \citenamefont {Roberts}, \citenamefont {Aranson},\ and\ \citenamefont
  {De~Pablo}}]{zhang2016}%
  \BibitemOpen
  \bibfield  {author} {\bibinfo {author} {\bibfnamefont {R.}~\bibnamefont
  {Zhang}}, \bibinfo {author} {\bibfnamefont {T.}~\bibnamefont {Roberts}},
  \bibinfo {author} {\bibfnamefont {I.~S.}\ \bibnamefont {Aranson}},\ and\
  \bibinfo {author} {\bibfnamefont {J.~J.}\ \bibnamefont {De~Pablo}},\
  }\bibfield  {title} {\bibinfo {title} {\emph {Lattice Boltzmann simulation of
  asymmetric flow in nematic liquid crystals with finite anchoring}},\ }\href
  {https://pubs.aip.org/aip/jcp/article/144/8/084905/912725} {\bibfield
  {journal} {\bibinfo  {journal} {\protect\JournalTitle{The Journal of Chemical
  Physics}}\ }\textbf {\bibinfo {volume} {144}} (\bibinfo {year}
  {2016})}\BibitemShut {NoStop}%
\bibitem [{\citenamefont {Beris}\ and\ \citenamefont
  {Edwards}(1994)}]{beris1994}%
  \BibitemOpen
  \bibfield  {author} {\bibinfo {author} {\bibfnamefont {A.~N.}\ \bibnamefont
  {Beris}}\ and\ \bibinfo {author} {\bibfnamefont {B.~J.}\ \bibnamefont
  {Edwards}},\ }\href@noop {} {\emph {\bibinfo {title} {Thermodynamics of
  flowing systems: with internal microstructure}}},\ \bibinfo {number} {36}\
  (\bibinfo  {publisher} {Oxford University Press, USA},\ \bibinfo {year}
  {1994})\BibitemShut {NoStop}%
\bibitem [{\citenamefont {Darnton}\ \emph {et~al.}(2004)\citenamefont
  {Darnton}, \citenamefont {Turner}, \citenamefont {Breuer},\ and\
  \citenamefont {Berg}}]{darnton2004}%
  \BibitemOpen
  \bibfield  {author} {\bibinfo {author} {\bibfnamefont {N.}~\bibnamefont
  {Darnton}}, \bibinfo {author} {\bibfnamefont {L.}~\bibnamefont {Turner}},
  \bibinfo {author} {\bibfnamefont {K.}~\bibnamefont {Breuer}},\ and\ \bibinfo
  {author} {\bibfnamefont {H.~C.}\ \bibnamefont {Berg}},\ }\bibfield  {title}
  {\bibinfo {title} {\emph {Moving fluid with bacterial carpets}},\ }\href
  {https://www.cell.com/AJHG/fulltext/S0006-3495(04)74253-8} {\bibfield
  {journal} {\bibinfo  {journal} {\protect\JournalTitle{Biophysical journal}}\
  }\textbf {\bibinfo {volume} {86}},\ \bibinfo {pages} {1863} (\bibinfo {year}
  {2004})}\BibitemShut {NoStop}%
\bibitem [{\citenamefont {Edwards}\ and\ \citenamefont
  {Yeomans}(2009)}]{edwards2009}%
  \BibitemOpen
  \bibfield  {author} {\bibinfo {author} {\bibfnamefont {S.}~\bibnamefont
  {Edwards}}\ and\ \bibinfo {author} {\bibfnamefont {J.}~\bibnamefont
  {Yeomans}},\ }\bibfield  {title} {\bibinfo {title} {\emph {Spontaneous flow
  states in active nematics: a unified picture}},\ }\href
  {https://doi.org/10.1209/0295-5075/85/18008} {\bibfield  {journal} {\bibinfo
  {journal} {\protect\JournalTitle{Europhysics letters}}\ }\textbf {\bibinfo
  {volume} {85}},\ \bibinfo {pages} {18008} (\bibinfo {year}
  {2009})}\BibitemShut {NoStop}%
\bibitem [{\citenamefont {Drescher}\ \emph {et~al.}(2010)\citenamefont
  {Drescher}, \citenamefont {Goldstein}, \citenamefont {Michel}, \citenamefont
  {Polin},\ and\ \citenamefont {Tuval}}]{drescher2010}%
  \BibitemOpen
  \bibfield  {author} {\bibinfo {author} {\bibfnamefont {K.}~\bibnamefont
  {Drescher}}, \bibinfo {author} {\bibfnamefont {R.~E.}\ \bibnamefont
  {Goldstein}}, \bibinfo {author} {\bibfnamefont {N.}~\bibnamefont {Michel}},
  \bibinfo {author} {\bibfnamefont {M.}~\bibnamefont {Polin}},\ and\ \bibinfo
  {author} {\bibfnamefont {I.}~\bibnamefont {Tuval}},\ }\bibfield  {title}
  {\bibinfo {title} {\emph {Direct measurement of the flow field around
  swimming microorganisms}},\ }\href
  {https://doi.org/10.1103/PhysRevLett.105.168101} {\bibfield  {journal}
  {\bibinfo  {journal} {\protect\JournalTitle{Physical Review Letters}}\
  }\textbf {\bibinfo {volume} {105}},\ \bibinfo {pages} {168101} (\bibinfo
  {year} {2010})}\BibitemShut {NoStop}%
\bibitem [{\citenamefont {Hintsche}\ \emph {et~al.}(2017)\citenamefont
  {Hintsche}, \citenamefont {Waljor}, \citenamefont {Gro{\ss}mann},
  \citenamefont {K{\"u}hn}, \citenamefont {Thormann}, \citenamefont {Peruani},\
  and\ \citenamefont {Beta}}]{hintsche2017}%
  \BibitemOpen
  \bibfield  {author} {\bibinfo {author} {\bibfnamefont {M.}~\bibnamefont
  {Hintsche}}, \bibinfo {author} {\bibfnamefont {V.}~\bibnamefont {Waljor}},
  \bibinfo {author} {\bibfnamefont {R.}~\bibnamefont {Gro{\ss}mann}}, \bibinfo
  {author} {\bibfnamefont {M.~J.}\ \bibnamefont {K{\"u}hn}}, \bibinfo {author}
  {\bibfnamefont {K.~M.}\ \bibnamefont {Thormann}}, \bibinfo {author}
  {\bibfnamefont {F.}~\bibnamefont {Peruani}},\ and\ \bibinfo {author}
  {\bibfnamefont {C.}~\bibnamefont {Beta}},\ }\bibfield  {title} {\bibinfo
  {title} {\emph {A polar bundle of flagella can drive bacterial swimming by
  pushing, pulling, or coiling around the cell body}},\ }\href
  {https://doi.org/https://doi.org/10.1038/s41598-017-16428-9} {\bibfield
  {journal} {\bibinfo  {journal} {\protect\JournalTitle{Scientific reports}}\
  }\textbf {\bibinfo {volume} {7}},\ \bibinfo {pages} {16771} (\bibinfo {year}
  {2017})}\BibitemShut {NoStop}%
\bibitem [{\citenamefont {Singh}\ \emph {et~al.}(2017)\citenamefont {Singh},
  \citenamefont {Hosseinidoust}, \citenamefont {Park}, \citenamefont {Yasa},\
  and\ \citenamefont {Sitti}}]{singh2017}%
  \BibitemOpen
  \bibfield  {author} {\bibinfo {author} {\bibfnamefont {A.~V.}\ \bibnamefont
  {Singh}}, \bibinfo {author} {\bibfnamefont {Z.}~\bibnamefont
  {Hosseinidoust}}, \bibinfo {author} {\bibfnamefont {B.-W.}\ \bibnamefont
  {Park}}, \bibinfo {author} {\bibfnamefont {O.}~\bibnamefont {Yasa}},\ and\
  \bibinfo {author} {\bibfnamefont {M.}~\bibnamefont {Sitti}},\ }\bibfield
  {title} {\bibinfo {title} {\emph {Microemulsion-based soft bacteria-driven
  microswimmers for active cargo delivery}},\ }\href
  {https://doi.org/10.1021/acsnano.7b02082} {\bibfield  {journal} {\bibinfo
  {journal} {\protect\JournalTitle{ACS nano}}\ }\textbf {\bibinfo {volume}
  {11}},\ \bibinfo {pages} {9759} (\bibinfo {year} {2017})}\BibitemShut
  {NoStop}%
\bibitem [{\citenamefont {Park}\ \emph {et~al.}(2017)\citenamefont {Park},
  \citenamefont {Zhuang}, \citenamefont {Yasa},\ and\ \citenamefont
  {Sitti}}]{park2017}%
  \BibitemOpen
  \bibfield  {author} {\bibinfo {author} {\bibfnamefont {B.-W.}\ \bibnamefont
  {Park}}, \bibinfo {author} {\bibfnamefont {J.}~\bibnamefont {Zhuang}},
  \bibinfo {author} {\bibfnamefont {O.}~\bibnamefont {Yasa}},\ and\ \bibinfo
  {author} {\bibfnamefont {M.}~\bibnamefont {Sitti}},\ }\bibfield  {title}
  {\bibinfo {title} {\emph {Multifunctional bacteria-driven microswimmers for
  targeted active drug delivery}},\ }\href@noop {} {\bibfield  {journal}
  {\bibinfo  {journal} {\protect\JournalTitle{ACS nano}}\ }\textbf {\bibinfo
  {volume} {11}},\ \bibinfo {pages} {8910} (\bibinfo {year}
  {2017})}\BibitemShut {NoStop}%
\bibitem [{\citenamefont {Alapan}\ \emph {et~al.}(2018)\citenamefont {Alapan},
  \citenamefont {Yasa}, \citenamefont {Schauer}, \citenamefont {Giltinan},
  \citenamefont {Tabak}, \citenamefont {Sourjik},\ and\ \citenamefont
  {Sitti}}]{alapan2018}%
  \BibitemOpen
  \bibfield  {author} {\bibinfo {author} {\bibfnamefont {Y.}~\bibnamefont
  {Alapan}}, \bibinfo {author} {\bibfnamefont {O.}~\bibnamefont {Yasa}},
  \bibinfo {author} {\bibfnamefont {O.}~\bibnamefont {Schauer}}, \bibinfo
  {author} {\bibfnamefont {J.}~\bibnamefont {Giltinan}}, \bibinfo {author}
  {\bibfnamefont {A.~F.}\ \bibnamefont {Tabak}}, \bibinfo {author}
  {\bibfnamefont {V.}~\bibnamefont {Sourjik}},\ and\ \bibinfo {author}
  {\bibfnamefont {M.}~\bibnamefont {Sitti}},\ }\bibfield  {title} {\bibinfo
  {title} {\emph {Soft erythrocyte-based bacterial microswimmers for cargo
  delivery}},\ }\href {https://doi.org/10.1126/scirobotics.aar4423} {\bibfield
  {journal} {\bibinfo  {journal} {\protect\JournalTitle{Science robotics}}\
  }\textbf {\bibinfo {volume} {3}},\ \bibinfo {pages} {eaar4423} (\bibinfo
  {year} {2018})}\BibitemShut {NoStop}%
\bibitem [{\citenamefont {Wu}\ \emph {et~al.}(2025)\citenamefont {Wu},
  \citenamefont {Liu}, \citenamefont {Mou}, \citenamefont {Li}, \citenamefont
  {Wang}, \citenamefont {Asilehan}, \citenamefont {Shi}, \citenamefont {You},
  \citenamefont {Zhang}, \citenamefont {Jiang},\ and\ \citenamefont
  {Peng}}]{Wu2025}%
  \BibitemOpen
  \bibfield  {author} {\bibinfo {author} {\bibfnamefont {J.}~\bibnamefont
  {Wu}}, \bibinfo {author} {\bibfnamefont {M.}~\bibnamefont {Liu}}, \bibinfo
  {author} {\bibfnamefont {Z.}~\bibnamefont {Mou}}, \bibinfo {author}
  {\bibfnamefont {Y.}~\bibnamefont {Li}}, \bibinfo {author} {\bibfnamefont
  {R.}~\bibnamefont {Wang}}, \bibinfo {author} {\bibfnamefont {Z.}~\bibnamefont
  {Asilehan}}, \bibinfo {author} {\bibfnamefont {Q.}~\bibnamefont {Shi}},
  \bibinfo {author} {\bibfnamefont {Z.}~\bibnamefont {You}}, \bibinfo {author}
  {\bibfnamefont {R.}~\bibnamefont {Zhang}}, \bibinfo {author} {\bibfnamefont
  {J.}~\bibnamefont {Jiang}},\ and\ \bibinfo {author} {\bibfnamefont
  {C.}~\bibnamefont {Peng}},\ }\bibfield  {title} {\bibinfo {title} {\emph
  {Programmable double traveling waves in living liquid crystals}}} (\bibinfo
  {year} {2025}),\ \bibinfo {note} {in preparation}\BibitemShut {NoStop}%
\bibitem [{\citenamefont {Ramaswamy}(2010)}]{ramaswamy2010}%
  \BibitemOpen
  \bibfield  {author} {\bibinfo {author} {\bibfnamefont {S.}~\bibnamefont
  {Ramaswamy}},\ }\bibfield  {title} {\bibinfo {title} {\emph {The mechanics
  and statistics of active matter}},\ }\href
  {https://doi.org/https://doi.org/10.1146/annurev-conmatphys-070909-104101}
  {\bibfield  {journal} {\bibinfo  {journal} {\protect\JournalTitle{Annu. Rev.
  Condens. Matter Phys.}}\ }\textbf {\bibinfo {volume} {1}},\ \bibinfo {pages}
  {323} (\bibinfo {year} {2010})}\BibitemShut {NoStop}%
\bibitem [{\citenamefont {Duclos}\ \emph {et~al.}(2020)\citenamefont {Duclos},
  \citenamefont {Adkins}, \citenamefont {Banerjee}, \citenamefont {Peterson},
  \citenamefont {Varghese}, \citenamefont {Kolvin}, \citenamefont {Baskaran},
  \citenamefont {Pelcovits}, \citenamefont {Powers}, \citenamefont {Baskaran}
  \emph {et~al.}}]{duclos2020}%
  \BibitemOpen
  \bibfield  {author} {\bibinfo {author} {\bibfnamefont {G.}~\bibnamefont
  {Duclos}}, \bibinfo {author} {\bibfnamefont {R.}~\bibnamefont {Adkins}},
  \bibinfo {author} {\bibfnamefont {D.}~\bibnamefont {Banerjee}}, \bibinfo
  {author} {\bibfnamefont {M.~S.}\ \bibnamefont {Peterson}}, \bibinfo {author}
  {\bibfnamefont {M.}~\bibnamefont {Varghese}}, \bibinfo {author}
  {\bibfnamefont {I.}~\bibnamefont {Kolvin}}, \bibinfo {author} {\bibfnamefont
  {A.}~\bibnamefont {Baskaran}}, \bibinfo {author} {\bibfnamefont {R.~A.}\
  \bibnamefont {Pelcovits}}, \bibinfo {author} {\bibfnamefont {T.~R.}\
  \bibnamefont {Powers}}, \bibinfo {author} {\bibfnamefont {A.}~\bibnamefont
  {Baskaran}}, \emph {et~al.},\ }\bibfield  {title} {\bibinfo {title} {\emph
  {Topological structure and dynamics of three-dimensional active nematics}},\
  }\href {https://doi.org/https://www.science.org/doi/10.1126/science.aaz4547}
  {\bibfield  {journal} {\bibinfo  {journal} {\protect\JournalTitle{Science}}\
  }\textbf {\bibinfo {volume} {367}},\ \bibinfo {pages} {1120} (\bibinfo {year}
  {2020})}\BibitemShut {NoStop}%
\bibitem [{\citenamefont {Binysh}\ \emph {et~al.}(2020)\citenamefont {Binysh},
  \citenamefont {Kos}, \citenamefont {\ifmmode~\check{C}\else \v{C}\fi{}opar},
  \citenamefont {Ravnik},\ and\ \citenamefont {Alexander}}]{binysh2020}%
  \BibitemOpen
  \bibfield  {author} {\bibinfo {author} {\bibfnamefont {J.}~\bibnamefont
  {Binysh}}, \bibinfo {author} {\bibfnamefont {i.~c.~v.}\ \bibnamefont {Kos}},
  \bibinfo {author} {\bibfnamefont {S.}~\bibnamefont {\ifmmode~\check{C}\else
  \v{C}\fi{}opar}}, \bibinfo {author} {\bibfnamefont {M.}~\bibnamefont
  {Ravnik}},\ and\ \bibinfo {author} {\bibfnamefont {G.~P.}\ \bibnamefont
  {Alexander}},\ }\bibfield  {title} {\bibinfo {title} {\emph
  {Three-Dimensional Active Defect Loops}},\ }\href
  {https://doi.org/10.1103/PhysRevLett.124.088001} {\bibfield  {journal}
  {\bibinfo  {journal} {\protect\JournalTitle{Phys. Rev. Lett.}}\ }\textbf
  {\bibinfo {volume} {124}},\ \bibinfo {pages} {088001} (\bibinfo {year}
  {2020})}\BibitemShut {NoStop}%
\bibitem [{\citenamefont {Long}\ \emph {et~al.}(2021)\citenamefont {Long},
  \citenamefont {Tang}, \citenamefont {Selinger},\ and\ \citenamefont
  {Selinger}}]{long2021}%
  \BibitemOpen
  \bibfield  {author} {\bibinfo {author} {\bibfnamefont {C.}~\bibnamefont
  {Long}}, \bibinfo {author} {\bibfnamefont {X.}~\bibnamefont {Tang}}, \bibinfo
  {author} {\bibfnamefont {R.~L.}\ \bibnamefont {Selinger}},\ and\ \bibinfo
  {author} {\bibfnamefont {J.~V.}\ \bibnamefont {Selinger}},\ }\bibfield
  {title} {\bibinfo {title} {\emph {Geometry and mechanics of disclination
  lines in 3D nematic liquid crystals}},\ }\href
  {https://doi.org/https://doi.org/10.1039/D0SM01899F} {\bibfield  {journal}
  {\bibinfo  {journal} {\protect\JournalTitle{Soft Matter}}\ }\textbf {\bibinfo
  {volume} {17}},\ \bibinfo {pages} {2265} (\bibinfo {year}
  {2021})}\BibitemShut {NoStop}%
\bibitem [{\citenamefont {Kralj}\ \emph {et~al.}(2023)\citenamefont {Kralj},
  \citenamefont {Ravnik},\ and\ \citenamefont {Kos}}]{kralj2023}%
  \BibitemOpen
  \bibfield  {author} {\bibinfo {author} {\bibfnamefont {N.}~\bibnamefont
  {Kralj}}, \bibinfo {author} {\bibfnamefont {M.}~\bibnamefont {Ravnik}},\ and\
  \bibinfo {author} {\bibfnamefont {i.~c.~v.}\ \bibnamefont {Kos}},\ }\bibfield
   {title} {\bibinfo {title} {\emph {Defect Line Coarsening and Refinement in
  Active Nematics}},\ }\href {https://doi.org/10.1103/PhysRevLett.130.128101}
  {\bibfield  {journal} {\bibinfo  {journal} {\protect\JournalTitle{Phys. Rev.
  Lett.}}\ }\textbf {\bibinfo {volume} {130}},\ \bibinfo {pages} {128101}
  (\bibinfo {year} {2023})}\BibitemShut {NoStop}%
\bibitem [{\citenamefont {Goral}\ \emph {et~al.}(2022)\citenamefont {Goral},
  \citenamefont {Clement}, \citenamefont {Darnige}, \citenamefont
  {Lopez-Leon},\ and\ \citenamefont {Lindner}}]{goral2022}%
  \BibitemOpen
  \bibfield  {author} {\bibinfo {author} {\bibfnamefont {M.}~\bibnamefont
  {Goral}}, \bibinfo {author} {\bibfnamefont {E.}~\bibnamefont {Clement}},
  \bibinfo {author} {\bibfnamefont {T.}~\bibnamefont {Darnige}}, \bibinfo
  {author} {\bibfnamefont {T.}~\bibnamefont {Lopez-Leon}},\ and\ \bibinfo
  {author} {\bibfnamefont {A.}~\bibnamefont {Lindner}},\ }\bibfield  {title}
  {\bibinfo {title} {\emph {Frustrated ‘run and tumble’of swimming
  Escherichia coli bacteria in nematic liquid crystals}},\ }\href
  {https://doi.org/https://doi.org/10.1098/rsfs.2022.0039} {\bibfield
  {journal} {\bibinfo  {journal} {\protect\JournalTitle{Interface Focus}}\
  }\textbf {\bibinfo {volume} {12}},\ \bibinfo {pages} {20220039} (\bibinfo
  {year} {2022})}\BibitemShut {NoStop}%
\bibitem [{\citenamefont {Prabhune}\ \emph {et~al.}(2024)\citenamefont
  {Prabhune}, \citenamefont {Garc\'{\i}a-Gordillo}, \citenamefont {Aranson},
  \citenamefont {Powers},\ and\ \citenamefont
  {Figueroa-Morales}}]{prabhune2024}%
  \BibitemOpen
  \bibfield  {author} {\bibinfo {author} {\bibfnamefont {A.~G.}\ \bibnamefont
  {Prabhune}}, \bibinfo {author} {\bibfnamefont {A.~S.}\ \bibnamefont
  {Garc\'{\i}a-Gordillo}}, \bibinfo {author} {\bibfnamefont {I.~S.}\
  \bibnamefont {Aranson}}, \bibinfo {author} {\bibfnamefont {T.~R.}\
  \bibnamefont {Powers}},\ and\ \bibinfo {author} {\bibfnamefont
  {N.}~\bibnamefont {Figueroa-Morales}},\ }\bibfield  {title} {\bibinfo {title}
  {\emph {Bacteria Navigate Anisotropic Media using a Flagellar Tug-of-Oars}},\
  }\href {https://doi.org/10.1103/PRXLife.2.033004} {\bibfield  {journal}
  {\bibinfo  {journal} {\protect\JournalTitle{PRX Life}}\ }\textbf {\bibinfo
  {volume} {2}},\ \bibinfo {pages} {033004} (\bibinfo {year}
  {2024})}\BibitemShut {NoStop}%
\bibitem [{\citenamefont {Berg}\ and\ \citenamefont {Brown}(1972)}]{berg1972}%
  \BibitemOpen
  \bibfield  {author} {\bibinfo {author} {\bibfnamefont {H.~C.}\ \bibnamefont
  {Berg}}\ and\ \bibinfo {author} {\bibfnamefont {D.~A.}\ \bibnamefont
  {Brown}},\ }\bibfield  {title} {\bibinfo {title} {\emph {Chemotaxis in
  Escherichia coli analysed by three-dimensional tracking}},\ }\href
  {https://doi.org/https://doi.org/10.1038/239500a0} {\bibfield  {journal}
  {\bibinfo  {journal} {\protect\JournalTitle{nature}}\ }\textbf {\bibinfo
  {volume} {239}},\ \bibinfo {pages} {500} (\bibinfo {year}
  {1972})}\BibitemShut {NoStop}%
\bibitem [{\citenamefont {Taute}\ \emph {et~al.}(2015)\citenamefont {Taute},
  \citenamefont {Gude}, \citenamefont {Tans},\ and\ \citenamefont
  {Shimizu}}]{taute2015}%
  \BibitemOpen
  \bibfield  {author} {\bibinfo {author} {\bibfnamefont {K.}~\bibnamefont
  {Taute}}, \bibinfo {author} {\bibfnamefont {S.}~\bibnamefont {Gude}},
  \bibinfo {author} {\bibfnamefont {S.}~\bibnamefont {Tans}},\ and\ \bibinfo
  {author} {\bibfnamefont {T.}~\bibnamefont {Shimizu}},\ }\bibfield  {title}
  {\bibinfo {title} {\emph {High-throughput 3D tracking of bacteria on a
  standard phase contrast microscope}},\ }\href
  {https://doi.org/https://doi.org/10.1038/ncomms9776} {\bibfield  {journal}
  {\bibinfo  {journal} {\protect\JournalTitle{Nature communications}}\ }\textbf
  {\bibinfo {volume} {6}},\ \bibinfo {pages} {1} (\bibinfo {year}
  {2015})}\BibitemShut {NoStop}%
\bibitem [{\citenamefont {Figueroa-Morales}\ \emph {et~al.}(2020)\citenamefont
  {Figueroa-Morales}, \citenamefont {Soto}, \citenamefont {Junot},
  \citenamefont {Darnige}, \citenamefont {Douarche}, \citenamefont {Martinez},
  \citenamefont {Lindner},\ and\ \citenamefont
  {Cl\'ement}}]{Figueroa-Morales2020}%
  \BibitemOpen
  \bibfield  {author} {\bibinfo {author} {\bibfnamefont {N.}~\bibnamefont
  {Figueroa-Morales}}, \bibinfo {author} {\bibfnamefont {R.}~\bibnamefont
  {Soto}}, \bibinfo {author} {\bibfnamefont {G.}~\bibnamefont {Junot}},
  \bibinfo {author} {\bibfnamefont {T.}~\bibnamefont {Darnige}}, \bibinfo
  {author} {\bibfnamefont {C.}~\bibnamefont {Douarche}}, \bibinfo {author}
  {\bibfnamefont {V.~A.}\ \bibnamefont {Martinez}}, \bibinfo {author}
  {\bibfnamefont {A.}~\bibnamefont {Lindner}},\ and\ \bibinfo {author}
  {\bibfnamefont {E.}~\bibnamefont {Cl\'ement}},\ }\bibfield  {title} {\bibinfo
  {title} {\emph {3D Spatial Exploration by E. coli Echoes Motor Temporal
  Variability}},\ }\href {https://doi.org/10.1103/PhysRevX.10.021004}
  {\bibfield  {journal} {\bibinfo  {journal} {\protect\JournalTitle{Phys. Rev.
  X}}\ }\textbf {\bibinfo {volume} {10}},\ \bibinfo {pages} {021004} (\bibinfo
  {year} {2020})}\BibitemShut {NoStop}%
\bibitem [{\citenamefont {Chan}\ \emph
  {et~al.}(2024{\natexlab{b}})\citenamefont {Chan}, \citenamefont {Yang},
  \citenamefont {Gan},\ and\ \citenamefont {Zhang}}]{chan2024interplay}%
  \BibitemOpen
  \bibfield  {author} {\bibinfo {author} {\bibfnamefont {C.~W.}\ \bibnamefont
  {Chan}}, \bibinfo {author} {\bibfnamefont {Z.}~\bibnamefont {Yang}}, \bibinfo
  {author} {\bibfnamefont {Z.}~\bibnamefont {Gan}},\ and\ \bibinfo {author}
  {\bibfnamefont {R.}~\bibnamefont {Zhang}},\ }\bibfield  {title} {\bibinfo
  {title} {\emph {Interplay of chemotactic force, P{\'e}clet number, and
  dimensionality dictates the dynamics of auto-chemotactic chiral active
  droplets}},\ }\bibfield  {journal} {\bibinfo  {journal}
  {\protect\JournalTitle{The Journal of Chemical Physics}}\ }\textbf {\bibinfo
  {volume} {161}},\ \href {https://doi.org/https://doi.org/10.1063/5.0207355}
  {https://doi.org/10.1063/5.0207355} (\bibinfo {year}
  {2024}{\natexlab{b}})\BibitemShut {NoStop}%
\bibitem [{\citenamefont {\ifmmode~\check{C}\else \v{C}\fi{}opar}\ \emph
  {et~al.}(2019)\citenamefont {\ifmmode~\check{C}\else \v{C}\fi{}opar},
  \citenamefont {Aplinc}, \citenamefont {Kos}, \citenamefont
  {\ifmmode~\check{Z}\else \v{Z}\fi{}umer},\ and\ \citenamefont
  {Ravnik}}]{copar2019}%
  \BibitemOpen
  \bibfield  {author} {\bibinfo {author} {\bibfnamefont {S.}~\bibnamefont
  {\ifmmode~\check{C}\else \v{C}\fi{}opar}}, \bibinfo {author} {\bibfnamefont
  {J.}~\bibnamefont {Aplinc}}, \bibinfo {author} {\bibfnamefont {i.~c.~v.}\
  \bibnamefont {Kos}}, \bibinfo {author} {\bibfnamefont {S.}~\bibnamefont
  {\ifmmode~\check{Z}\else \v{Z}\fi{}umer}},\ and\ \bibinfo {author}
  {\bibfnamefont {M.}~\bibnamefont {Ravnik}},\ }\bibfield  {title} {\bibinfo
  {title} {\emph {Topology of Three-Dimensional Active Nematic Turbulence
  Confined to Droplets}},\ }\href {https://doi.org/10.1103/PhysRevX.9.031051}
  {\bibfield  {journal} {\bibinfo  {journal} {\protect\JournalTitle{Phys. Rev.
  X}}\ }\textbf {\bibinfo {volume} {9}},\ \bibinfo {pages} {031051} (\bibinfo
  {year} {2019})}\BibitemShut {NoStop}%
\bibitem [{\citenamefont {Carenza}\ \emph {et~al.}(2019)\citenamefont
  {Carenza}, \citenamefont {Gonnella}, \citenamefont {Marenduzzo},\ and\
  \citenamefont {Negro}}]{carenza2019}%
  \BibitemOpen
  \bibfield  {author} {\bibinfo {author} {\bibfnamefont {L.~N.}\ \bibnamefont
  {Carenza}}, \bibinfo {author} {\bibfnamefont {G.}~\bibnamefont {Gonnella}},
  \bibinfo {author} {\bibfnamefont {D.}~\bibnamefont {Marenduzzo}},\ and\
  \bibinfo {author} {\bibfnamefont {G.}~\bibnamefont {Negro}},\ }\bibfield
  {title} {\bibinfo {title} {\emph {Rotation and propulsion in 3D active chiral
  droplets}},\ }\href {https://doi.org/https://doi.org/10.1073/pnas.1910909116}
  {\bibfield  {journal} {\bibinfo  {journal} {\protect\JournalTitle{Proceedings
  of the National Academy of Sciences}}\ }\textbf {\bibinfo {volume} {116}},\
  \bibinfo {pages} {22065} (\bibinfo {year} {2019})}\BibitemShut {NoStop}%
\bibitem [{\citenamefont {Whitfield}\ \emph {et~al.}(2017)\citenamefont
  {Whitfield}, \citenamefont {Adhyapak}, \citenamefont {Tiribocchi},
  \citenamefont {Alexander}, \citenamefont {Marenduzzo},\ and\ \citenamefont
  {Ramaswamy}}]{whitfield2017}%
  \BibitemOpen
  \bibfield  {author} {\bibinfo {author} {\bibfnamefont {C.~A.}\ \bibnamefont
  {Whitfield}}, \bibinfo {author} {\bibfnamefont {T.~C.}\ \bibnamefont
  {Adhyapak}}, \bibinfo {author} {\bibfnamefont {A.}~\bibnamefont
  {Tiribocchi}}, \bibinfo {author} {\bibfnamefont {G.~P.}\ \bibnamefont
  {Alexander}}, \bibinfo {author} {\bibfnamefont {D.}~\bibnamefont
  {Marenduzzo}},\ and\ \bibinfo {author} {\bibfnamefont {S.}~\bibnamefont
  {Ramaswamy}},\ }\bibfield  {title} {\bibinfo {title} {\emph {Hydrodynamic
  instabilities in active cholesteric liquid crystals}},\ }\href
  {https://doi.org/https://doi.org/10.1140/epje/i2017-11536-2} {\bibfield
  {journal} {\bibinfo  {journal} {\protect\JournalTitle{The European Physical
  Journal E}}\ }\textbf {\bibinfo {volume} {40}},\ \bibinfo {pages} {1}
  (\bibinfo {year} {2017})}\BibitemShut {NoStop}%
\bibitem [{\citenamefont {Boule}\ \emph {et~al.}(2020)\citenamefont {Boule},
  \citenamefont {Rainville},\ and\ \citenamefont {Galstian}}]{boule2020}%
  \BibitemOpen
  \bibfield  {author} {\bibinfo {author} {\bibfnamefont {M.-A.}\ \bibnamefont
  {Boule}}, \bibinfo {author} {\bibfnamefont {S.}~\bibnamefont {Rainville}},\
  and\ \bibinfo {author} {\bibfnamefont {T.}~\bibnamefont {Galstian}},\
  }\bibfield  {title} {\bibinfo {title} {\emph {Dynamic guiding of bacteria in
  lyotropic chromonic liquid crystal using magnetic field}},\ }\href
  {https://doi.org/https://www.tandfonline.com/doi/abs/10.1080/15421406.2020.1848229}
  {\bibfield  {journal} {\bibinfo  {journal} {\protect\JournalTitle{Molecular
  Crystals and Liquid Crystals}}\ }\textbf {\bibinfo {volume} {712}},\ \bibinfo
  {pages} {10} (\bibinfo {year} {2020})}\BibitemShut {NoStop}%
\end{thebibliography}%

\end{document}